\documentclass{aa}  
\newdimen\imgwidth
\usepackage{graphicx}
\usepackage[usenames]{color}
\usepackage{longtable}
\usepackage{dcolumn}

\usepackage{amsmath}
\usepackage{amsfonts}
\usepackage{supertabular}
\usepackage{txfonts}

\makeatletter
\newlength{\VSpaceBeforeTabBib}
\newlength{\VSpaceBeforeTabFoot}
\newcommand*\aa@tablefootname{Note}
\newcommand*\aa@tablefootfont{\small}
\newcommand*\aa@tablefootnamefont{\small\bfseries}
\newcommand\tablefoot[1]{\VSpaceBeforeTabBib=1ex%
  \par\vspace{\VSpaceBeforeTabFoot}
  \noindent
  \begin{minipage}{\linewidth}
    {\aa@tablefootnamefont\aa@tablefootname.}~%
    \aa@tablefootfont
    \ignorespaces
    #1%
  \end{minipage}%
}
\makeatother

\begin{document}
   \title{The shape of broad-line profiles in AGN} 
%\thanks{Based on Sloan data}
% }

   \author{W. Kollatschny 
          ,  M. Zetzl 
          }

   \institute{Institut f\"ur Astrophysik, Universit\"at G\"ottingen,
              Friedrich-Hund Platz 1, D-37077 G\"ottingen, Germany\\
              \email{wkollat@astro.physik.uni-goettingen.de}
}

%   \offprints{???}

   \date{Received 2012; accepted 2012}
   \authorrunning{Kollatschny, Zetzl}
   \titlerunning{Shape of AGN line profiles}
% \abstract{}{}{}{}{} 
% 5 {} token are mandatory

%$^{\star\star\star}$.

 \abstract{}
{We present
a study of the broad optical/UV emission line profiles in AGN
(active galactic nuclei) to get information on
the dominant velocity components (turbulence, rotation, etc.)
 in the central broad-line region (BLR).}
{We introduce line broadening simulations
 of emission line profiles and compare these results
with the largest homogeneous data set of reverberation-mapped AGN.}
{The underlying 
broad-line profiles in AGN are
Lorentzian profiles caused by turbulence in the line emitting region.
The turbulent velocities are different for the different line emitting regions
of H$\gamma$, H$\alpha$,
Ly$\alpha$, \ion{C}{iii}]\,$\lambda 1909$, \ion{He}{ii}\,$\lambda 1640$,
and \ion{Si}{iv}\,$\lambda 1400$.
The turbulent velocities go from 400\,$km\,s^{-1}$  for H$\beta$ up to
 3,800\,$km\,s^{-1}$ for Ly$\alpha$+\ion{N}{v}\,$\lambda 1240$.
The dominating broadening mechanism of these profiles
is broadening due to rotation. The rotation velocities
causing the line profile broadening 
 go from 500\,$km\,s^{-1}$ up to 6,500\,$km\,s^{-1}$.
Here we present
interrelations between observed emission line widths (FWHM)
and their related rotational velocities to correct
for the contribution of the turbulence to the broad-line profiles.
}
{}


  \maketitle
%
%________________________________________________________________
  
\section{Introduction}
It is now generally accepted that a super-massive black hole 
in the center of active galactic nuclei (AGN) is surrounded by an
accretion disk (e.g. Urry \& Padovani \cite{urry95}). In the outer regionsof the accretion disk, the broad emission lines are created by
photoionization. However, many details of this line emitting region are unknown,
and there are many models that treat their geometry and structure
(e.g. Collin-Souffrin et al. \cite{collin88},
 Emmering et al. \cite{emmering92},  
 K\"onigl \& Kartje \cite{koenigl94}, Murray \& Chiang  \cite{murray97}, 
Elvis \cite{elvis00}, Kollatschny \cite{kollatschny03a}, Ho \cite{ho08},
Gaskell \cite{gaskell10}, Goad et al. \cite{goad12} and references therein).
This broad-line region is spatially not resolved on direct images.

Information on the structure
and kinematics of this region can in principle be obtained
 from the broad emission line profiles.
However, there is the ambiguity in the possibility
that different geometries and different kinematics 
result in emission line profiles that have similar shapes.
This problem makes it difficult
to disentangle the profile contribution of individual parameters
(e.g. Netzer \cite{netzer90}, 
  Sulentic et al. \cite{sulentic00},
Zamfir et al. \cite{zamfir10}, and references therein).
 The shape and the width of the emission line profiles in AGN
might depend, among others parameters, on the velocity field,
on the geometry of the line emitting gas, on obscuration effects,
on the superposition of line emission from different regions,
and on the isotropy/anisotropy of the line emission. 

The velocity field might be a superposition of different components, such as
Doppler motions, turbulence, shock components, in/outflow components,
and rotation. 
Different velocity components result in different profiles, and the final
profile is a convolution of different components.
Doppler motions in the gravitational potential of a central black hole can
result in Gaussian profiles, where the width of the lines reflects
the Doppler motion of the line emitting gas
(Peterson \& Wandel \cite{peterson99}, Fromerth \& Melia \cite{fromerth00},
Peterson \& Wandel, \cite{peterson00}).
Logarithmic profiles agree well with observations of emission line wings
and can be produced by outflowing gas
%These logarithmic profiles agree well with observed profiles except
%in the cores as well as in the outer wings of the profiles 
(e.g.  Blumenthal \& Mathews \cite{blumenthal75}, Netzer \cite{netzer90}). 
The existence of Lorentzian profiles is consistent with emission
from extended accretion disks (Veron-Cetty et al. \cite{veron01},
Sulentic et al. \cite{sulentic02}).
Goad et al. (\cite{goad12})
demonstrate that turbulent motions in the outer
accretion disk produce Lorentzian profiles.
Furthermore, Lorentzian profiles might be explained by shock breakouts
 in dense winds (Moriya \& Tominaga \cite{moriya12}).
%Woertlich: A single Lorentzian profile represents emissiom from
%a classical BLR of
%fast-moving clouds. There are two sets of models including accretion discs:
%one set has the accretion disk plus a Lorentzian profile to represent a
%standard BLR, while the other set includes broad emission from the
%accretion disk only. Emission from the BLRs was modeled as a Lorentzian line,
%which is collisionally broadened with a width proportional to P/T, where
%P is the pressure and T the temperature. Down et al., 2010, MNRAS 401, 633.
%The importance of turbulence for the stability of accretion disks has
%been emphasized Osterbrock \cite{osterbrock78}).
In addition to these base-line profiles, both rotation and electron
scattering (e.g. H$\alpha$ in NGC~4593; Laor \cite{laor06}) can lead
to broadening of the line profiles.

Osterbrock pointed out as early as 
\cite{osterbrock78} that a combination of turbulence with rotation
 agrees well with the distribution of observed line widths of Seyfert~1
galaxies known at that time.
More recently, researchers have explored the possibility of an
additional net infall component (Gaskell \cite{gaskell10} for a review,
Hu et al. \cite{hu12};
but see Sulentic et al. \cite{sulentic12}) and an additional outflow
component in luminous AGNs (e.g. Wang et al. \cite{wang11}).
%In a recent contribution Gaskell (\cite{gaskell10})
%reviewed the evidence for an additional net infall component
%of the broad-line region from velocity-resolved reverberation
%mapping studies. On the other hand Sulentic et al. (\cite{sulentic12}) found
%no evidence for infall in a detailed
%study on the dynamics in the FeII emission region.
%Furthermore, there exist some evidence for an additional
%outflow component in luminous AGN (e.g. Wang et al. \cite{wang11}). 

The contribution of the velocity components might
be different from line to line,
as well as from galaxy to galaxy.
In their study of AGN line profiles
Robinson et al. (\cite{robinson95}) demonstrated that different types
of models produce similar kinds of profiles and that the profile shapes
differ between the AGN populations.
In addition, the optical and UV 
emission line profiles are similarly diverse in appearance.
In an investigation of broad emission lines from the Sloan Digital Sky Survey
(SDSS) Richards et al. (\cite{richards02}) point out
that the explanation
of their profiles is one of the most important tasks of future AGN models.
Using CIV emission line profiles, Richards et al. (\cite{richards11})
show that
the differences in the accretion disk wind between quasars can account
for some of the diversity of broad emission line profiles.

We demonstrated 
in a first paper (Kollatschny \& Zetzl \cite{kollatschny11},
hereafter called Paper I) that there is a general trend between 
the full-width at half maximum (FWHM) and the line width ratio 
FWHM/$\sigma_{line}$ in the broad emission lines of AGN.
Different emission lines exhibit different systematics in the 
FWHM/$\sigma_{line}$-vs-FWHM diagram. 
 The line width FWHM reflects the
rotational motion of the broad-line gas in combination with the associated
turbulent motion. This turbulent velocity is different for the
different emission lines.

Here we present a further study of
additional broad optical and UV emission line profiles of
reverberation-mapped AGN. The line profiles are parameterized by the ratio of
their full-width at half maximum (FWHM) to their line dispersion
$\sigma_{line}$. 

\section{Data: The line profile sample}

We studied in detail the mean profiles, as well as the
root-mean-square (rms) profiles of the
H$\beta$, \ion{He}{ii}\,$\lambda 4686$, and \ion{C}{iv}\,$\lambda 1550$  
lines of an AGN sample in Paper I.
This set of observations is the
largest homogeneous sample of variable AGN at this time, and it is based
on reverberation-mapped
AGN spectra (Peterson et al., \cite{peterson04}). 
The optical spectra were collected with different ground based telescopes while
the UV spectra were taken with the IUE satellite and the Hubble Space
Telescope. The sample has the advantage that all the spectra were reduced
in exactly the same way and that all spectra of each
galaxy were inter-calibrated with respect to each other.
The original AGN sample consists of 37 objects. 
In some cases multiple line spectra exist for a particular galaxy
based on different variability campaigns of the same galaxy. 

The present study is based on this data set as well.
Here we analyze all the remaining optical and UV broad-line profiles
we did not consider in Paper I, i.e., the
H$\gamma$, H$\alpha$, Ly$\alpha$, \ion{He}{ii}\,$\lambda 1640$,
\ion{C}{iii}]\,$\lambda 1909$, and
\ion{S}{iv}\,$\lambda 1400$ lines.

Usually the observed broad-line profiles in AGN are more or less contaminated by
additional narrow emission line components from the narrow line region.
For avoiding any major ambiguity we only inspected the
root-mean-square (rms) line profiles out of the sample of Peterson
 (\cite{peterson04}) as we did in Paper I.
 The rms profiles display the clean profiles of
 the variable broad emission lines.
 The narrow line components disappear in these
 spectra because
 they are constant on time scales of years. The optical and UV
 emission line profiles of our AGN sample are
 parameterized by their line widths (FWHM, and $\sigma_{line}$).
 The relationship between
FWHM and $\sigma_{line}$ contains 
information on the shape of the profile.
We use
all those line profiles of the sample that were regarded as
both reliable and less so by
Peterson (\cite{peterson04}) - same as in Paper I. 
%evtl etwas zu: The spectra cover the spectral range
% 3800 $\le \lambda \le$ 9200 \AA\ with a
%spectral resolution of $\lambda / \delta\lambda$  1800. 

All spectral data we are using in this current investigation are listed
in Table 1.
We present
H$\gamma${}, H$\alpha${}, Ly$\alpha${}, \ion{C}{iii}]\,$\lambda 1909$,
    \ion{Si}{iv}\,$\lambda 1400$, and \ion{He}{ii}\,$\lambda 1640$ line
widths (FWHM) and line width 
ratios (FWHM/$\sigma$) of all galaxies
in the AGN sample.

\newcommand{\Ha}{H$\alpha$}
\newcommand{\Hb}{H$\beta$}
\newcommand{\Hg}{H$\gamma$}
\newcommand{\Lya}{Ly$\alpha$}
\newcommand{\sVl}[3]{#1\,{\sc #2}]\,$\lambda{#3}$}
\newcommand{\Nl}[3]{#1\,{\sc #2}\,$\lambda{#3}$}
\makeatletter
 \def\hlinewd#1{%
   \noalign{\ifnum0=`}\fi\hrule \@height #1 \futurelet
    \reserved@a\@xhline}
\makeatother
\newcommand{\htopline}{\hlinewd{.8pt}}
\newcommand{\hmidline}{\hlinewd{.2pt}}
\newcommand{\hbotline}{\htopline}
\newcommand{\mcc}[1]{\multicolumn{1}{c}{#1}}
\newcommand{\mcl}[1]{\multicolumn{1}{l}{#1}}
\newcommand{\kms}{km\,s$^{-1}$}
\newcolumntype{p}{D{+}{\,\pm\,}{-1}}

\begin{table*}[htbp]
    \centering
       \leavevmode
       \tabcolsep8mm
        \newcolumntype{d}{D{.}{.}{-2}} 
\caption{\Hg{}, \Ha{}, \Lya{}, \sVl{C}{iii}{1909},
    \Nl{Si}{iv}{1400}, and \Nl{He}{ii}{1640} line widths of our AGN sample.} 
\begin{tabular}{llppp}
 \htopline
 \mcl{Object} & \mcl{Line} & \mcc{FWHM}& \mcc{$\sigma$} &  \mcc{FWHM/$\sigma$}\\
  &  & \mcc{[\kms{}]}& \mcc{ [\kms{}]} & \\
 \hmidline

PG1229+204 & \Hg{} & 3207 + 875 & 1540 + 159 & 2.082 + 0.169\\
PG0844+349 & \Hg{} & 4946 + 1085 & 2058 + 218 & 2.403 + 0.195\\
PG0052+251 & \Hg{} & 5633 + 3585 & 2230 + 502 & 2.526 + 0.432\\
PG0953+414 & \Hg{} & 2960 + 589 & 1299 + 193 & 2.279 + 0.261\\
PG1307+085 & \Hg{} & 4278 + 881 & 1758 + 193 & 2.433 + 0.204\\
PG2130+099 & \Hg{} & 2661 + 481 & 1836 + 191 & 1.449 + 0.130\\
PG1226+023 & \Hg{} & 3274 + 484 & 1688 + 142 & 1.940 + 0.130\\
PG0804+761: & \Hg{} & 2201 + 295 & 1256 + 141 & 1.752 + 0.160\\
PG1211+143: & \Hg{} & 2014 + 249 & 1376 + 157 & 1.464 + 0.143\\
\hmidline
PG0844+349 & \Ha{} & 2436 + 329 & 1625 + 73 & 1.499 + 0.057\\
PG1426+015 & \Ha{} & 5450 + 842 & 4254 + 290 & 1.281 + 0.078\\
PG1229+204 & \Ha{} & 3229 + 364 & 1737 + 118 & 1.859 + 0.101\\
PG1617+175 & \Ha{} & 3794 + 780 & 2483 + 160 & 1.528 + 0.083\\
PG1411+442 & \Ha{} & 1877 + 375 & 2437 + 196 & 0.770 + 0.072\\
PG0026+129 & \Ha{} & 1117 + 109 & 1961 + 135 & 0.570 + 0.056\\
PG0804+761 & \Ha{} & 3155 + 569 & 2046 + 138 & 1.542 + 0.088\\
PG2130+099 & \Ha{} & 1574 + 438 & 1421 + 80 & 1.108 + 0.059\\
NGC4151     & \Ha{} & 3156 + 300 & 2422 + 79 & 1.303 + 0.038\\
NGC4593     & \Ha{} & 3399 + 196 & 1253 + 90 & 2.713 + 0.147\\
NGC7469     & \Ha{} & 1615 + 119 & 1164 + 68 & 1.387 + 0.071\\
NGC3516     & \Ha{} & 4770 + 893 & 2108 + 69 & 2.263 + 0.057\\
NGC3227     & \Ha{} & 3168 + 67 & 1977 + 134 & 1.602 + 0.091\\
PG0052+251: & \Ha{} & 2682 + 453 & 1913 + 85 & 1.402 + 0.054\\
NGC4151:    & \Ha{} & 3724 + 529 & 1721 + 47 & 2.164 + 0.046\\
PG1211+143: & \Ha{} & 1425 + 382 & 2321 + 231 & 0.614 + 0.083\\
PG1226+023: & \Ha{} & 1638 + 424 & 2075 + 239 & 0.789 + 0.104\\
PG1307+085: & \Ha{} & 3084 + 1041 & 1843 + 98 & 1.673 + 0.073\\
Mrk279:     & \Ha{} & 3408 + 555 & 1405 + 266 & 2.426 + 0.351\\
NGC5548:    & \Ha{} & 3044 + 381 & 1694 + 80 & 1.797 + 0.069\\
\hmidline
Fairall9    & \Lya & 3503 + 1474 & 4120 + 308 & 0.850 + 0.069\\
3C390.3     & \Lya & 8732 + 985 & 3952 + 203 & 2.210 + 0.088\\
3C390.3:    & \Lya & 8225 + 781 & 4600 + 141 & 1.788 + 0.044\\
\hmidline
NGC5548 &\sVl{C}{iii}{1909}  & 4895 + 1263 & 3227 + 176 & 1.517 + 0.070\\
NGC5548 &\sVl{C}{iii}{1909}  & 5018 + 1458 & 2360 + 222 & 2.126 + 0.156\\
 \hmidline
NGC7469 & \Nl{Si}{iv}{1400} & 6033 + 1112 & 3495 + 269 & 1.726 + 0.109\\
NGC3783 & \Nl{Si}{iv}{1400} & 6343 + 2021 & 3488 + 161 & 1.819 + 0.068\\
NGC5548 & \Nl{Si}{iv}{1400} & 7044 + 1849 & 4014 + 253 & 1.755 + 0.090\\
NGC5548 & \Nl{Si}{iv}{1400} & 6455 + 3030 & 2576 + 389 & 2.506 + 0.288\\
\hmidline
NGC7469 & \Nl{He}{ii}{1640} & 10725 + 1697 & 3723 + 113 & 2.881 + 0.065\\
NGC3783 & \Nl{He}{ii}{1640} & 8008 + 1268 & 3870 + 162 & 2.069 + 0.068\\
NGC5548 & \Nl{He}{ii}{1640} & 8929 + 1571 & 4397 + 154 & 2.031 + 0.056\\
NGC5548 & \Nl{He}{ii}{1640} & 9803 + 1594 & 3897 + 264 & 2.516 + 0.130\\
\hmidline
%\end{supertabular}
\end{tabular}
\tablefoot{ Measurements of objects with colons (:) were
considered as less reliable by the authors
(Peterson et al., \cite{peterson04}).}
\end{table*}

\section{Results}

The following are the
emission line profiles resulting from various kinematical and
dynamical models 
that have been discussed in the context of the BLR in AGNs:\\
- Gaussian profiles due to Doppler motions:\\
\hspace*{10mm}  FWHM/$\sigma_{line} = 2.35$; \\
- Lorentzian profiles due to turbulent motions:\\
\hspace*{10mm}  FWHM/$\sigma_{line}  \approx 1~ (\rightarrow 0$); \\
- Exponential profiles caused by electron scattering:\\
\hspace*{10mm}  FWHM/$\sigma_{line} = \sqrt{2}\ln{2} \approx 0.98$;\\
- Logarithmic profiles caused by in-/outflow motions:\\
%\hspace*{10mm}  {\bf(FWHM/$\sigma_{line} \rightarrow 0$)}\\
\hspace*{10mm}  FWHM/$\sigma_{line} \approx 1~ (\rightarrow 0$). \\
We present in Fig. 1 all these emission line profiles scaled to the same
H$\beta$ line width (FWHM).
The exact mathematical value of FWHM/$\sigma_{line}$ goes to zero for
 Lorentzian profiles because of the infinitely extended wings
(see Fig. 1). However, the typical FWHM/$\sigma_{line}$ values we are measuring
are on the
order of one when integrating the spectral lines over a few hundred Angstroms.
Similarly the exact FWHM/$\sigma_{line}$ value for logarithmic profiles
cannot be calculated  because the central intensity goes to infinity.
See also discussions of this ratio in
%The FWHM/$\sigma_{line}$ ratios have been discussed by e.g. 
Peterson et al. (\cite{peterson04}), Collin et al. (\cite{collin06}), and
Goad et al. (\cite{goad12}).

In Fig. 2 we again show the basic logarithmic H$\beta$ profile for
$v_{min}$/$v_{max}$  = 0.003, as well as more line profiles 
generated by
expanding spherical shells of radiatively-driven isotropically emitting
clouds with different  $v_{min}$ /$v_{max}$ values ranging from 0.01
up to 0.9999 (see also Capriotti et al. \cite{capriotti80}).

\subsection{Line profile broadening simulations}

We are trying to find appropriate answers to the following questions
with our present investigation of AGN emission line profiles: 
which line profile is emitted intrinsically
and what broadening mechanisms
(besides the instrumental broadening) have an impact on the observed profile.
Lorentzian emission line profiles and/or Gaussian profiles are the most
accepted profiles that are thought to be emitted intrinsically.

We showed in Paper I that
rotational line broadening is indeed the most important broadening parameter
for AGN emission line profiles.  
We calculated the rotational line broadening by the
convolution of Lorentzian or Gaussian profiles with ellipsoidal profiles.
The rotational velocity $b =  \Delta\lambda/x$ is by definition the 
half width at zero intensity (HWZI) of an ellipsoidal profile    

\begin{equation}
\label{eq:Rot}
 A(x) = \frac{2}{\pi} \sqrt{1-x^{2}}. 
\end{equation}

In Paper I we used a numerical code that was developed by 
Hubeny et al. (\cite{hubeny94}) to compute the
line broadening due to rotation. This program 
considers limb darkening as well.
In the present study we use our own routine without limb darkening
that is based on the following broadening formula:\\

\begin{equation}
\label{eq:Conv}
 S(y) = \int\limits_{-\infty}^{+\infty}\!W(y-x)A(x)\,\text{d}x
\end{equation}\\
(Unsoeld, \cite{unsoeld55}), where  W is the intrinsic line profile without
rotational broadening, A the rotational profile, 
and S the convolved profile.

Other intrinsic line profiles and other broadening mechanisms than rotation
lead to different profile shapes. We computed
these profile shapes as well and compared them with observed
AGN profiles.

Figures 3 and 4
 show computed Lorentzian profiles that were
broadened by rotational motions. The rotation velocities range from
100\,$km\,s^{-1}$ to 8,000\,$km\,s^{-1}$. For the intrinsic
H$\beta$ Lorentzian profile
we adopted a turbulent velocity 
of 500\,$km\,s^{-1}$  (see Paper I) in the line emitting
region (Fig.~3).   
For the intrinsic Lorentzian profile of the \ion{C}{iv}\,$\lambda 1549$
line we adopted a turbulent velocity of the line emitting
region of 3,000\,$km\,s^{-1}$ (Fig.~4). 
We present additional line broadened profiles in Figs.~5 and 6. 
In Fig.~5 we made the assumption that the intrinsic H$\beta$ profile is
a Gaussian profile (FWHM = 500\,$km\,s^{-1}$) that is broadened by
rotation as in Fig.~3. Figure 6 demonstrates the line broadening of a 
Lorentzian H$\beta$ profile owing to Doppler motions
(convolution with a Gaussian profile). In all cases
we made the assumption that the integrated line intensities
remain constant.

\subsection{Observed and modeled line width ratios
 FWHM/ $\sigma_{line}$ versus line width FWHM. }

In the next step we present a  comparison of the observed line profile shapes
in different AGN spectra with computed line profiles.

\subsubsection{Rotational line broadening of Lorentzian H$\beta$, 
\ion{He}{ii}\,$\lambda 4686$,
\ion{C}{iv}\,$\lambda 1549$
 profiles}

In Paper I (their Figs. 1-3) we showed the observed H$\beta$,
\ion{He}{ii}\,$\lambda 4686$, 
and \ion{C}{iv}\,$\lambda 1549$ line width ratios
 FWHM/$\sigma_{line}$ versus their line width FWHM in AGN.
These line width data were taken from the sample of Peterson
et al. (\cite{peterson04}). 
We likewise included those measurements that
were regarded as less reliable by the authors
because they follow exactly the same trend.

We modeled the observations by rotationally broadened 
Lorentzian profiles.
We present in Fig. 7 the observed ratios 
 FWHM/$\sigma_{line}$ vs. FWHM
for the H$\beta$, \ion{He}{ii}\,$\lambda 4686$, and           
 \ion{C}{iv}\,$\lambda 1549$ profiles
in one single plot, as well as their modeling
with our new routine.
The rotational velocities in this figure are slightly shifted towards
higher FWHM (in comparison to Paper I) because the effect of limb darkening 
leads to a slightly modified rotational profile (see Fig. 168 in
Unsoeld, \cite{unsoeld55}).
The data for
these three line profiles and their corresponding models
fill different areas in this plot.
The underlying H$\beta$ Lorentzian profile has a line width of 
400\,$km\,s^{-1}$, the \ion{He}{ii}\,$\lambda 4686$ has an intrinsic
line width of 900\,$km\,s^{-1}$, and the \ion{C}{iv}\,$\lambda 1549$ line
has an intrinsic
line width of 2,900\,$km\,s^{-1}$. 
These line widths are least square fits to the data in Fig. 7.\\
The modeled rotational velocities of the 
H$\beta$ lines range from  
\renewcommand{\textfraction}{0}
%------------------------------------------------------------------------------
%
\onecolumn
\begin{figure}
\begin{minipage}[t]{0.475\textwidth}
\includegraphics[width=5.5cm,angle=270]{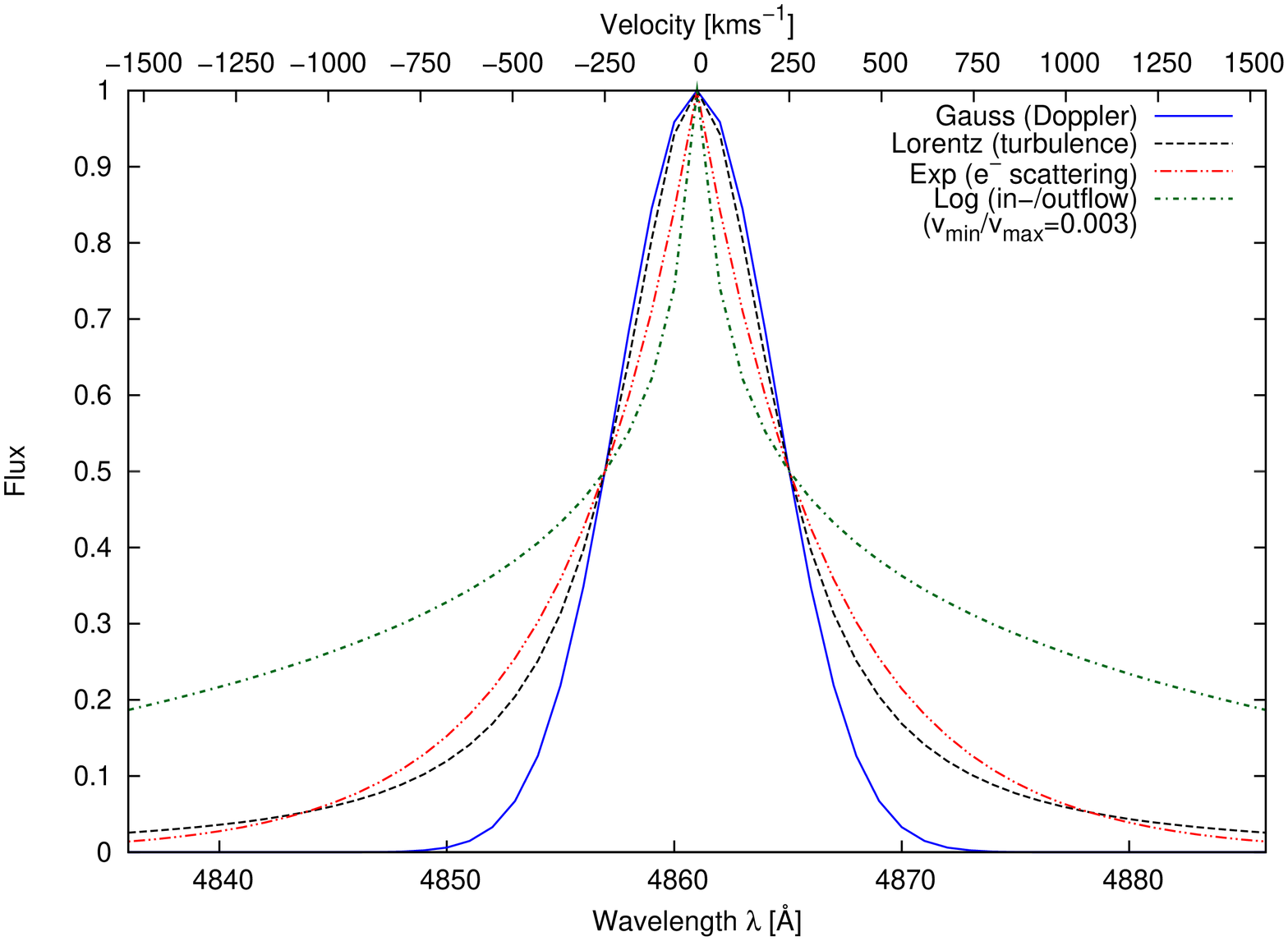} 
%\includegraphics[width=8.0cm,angle=0]{actdegrneigh.eps}
%\includegraphics[bb=40 90 380 700,width=63mm,angle=0]{actdegrneigh.eps}
%       \vspace*{5mm} 
%%       \vspace*{-2mm} 
  \caption{
Basic emission line profiles resulting from different kinematical models 
for the BLR in AGN:
Gaussian (solid line, blue), 
Lorentzian (dashed, black),
exponential (dot dot dashed, red), and
logarithmic (dot dashed, green) profiles.
All profiles are scaled to the same
H$\beta$ line width (FWHM=500\,$km\,s^{-1}$).}
   \label{profile_compare2.ps}
\end{minipage}
\hfill
\begin{minipage}[t]{0.475\textwidth}%
\includegraphics[width=5.5cm,angle=270]{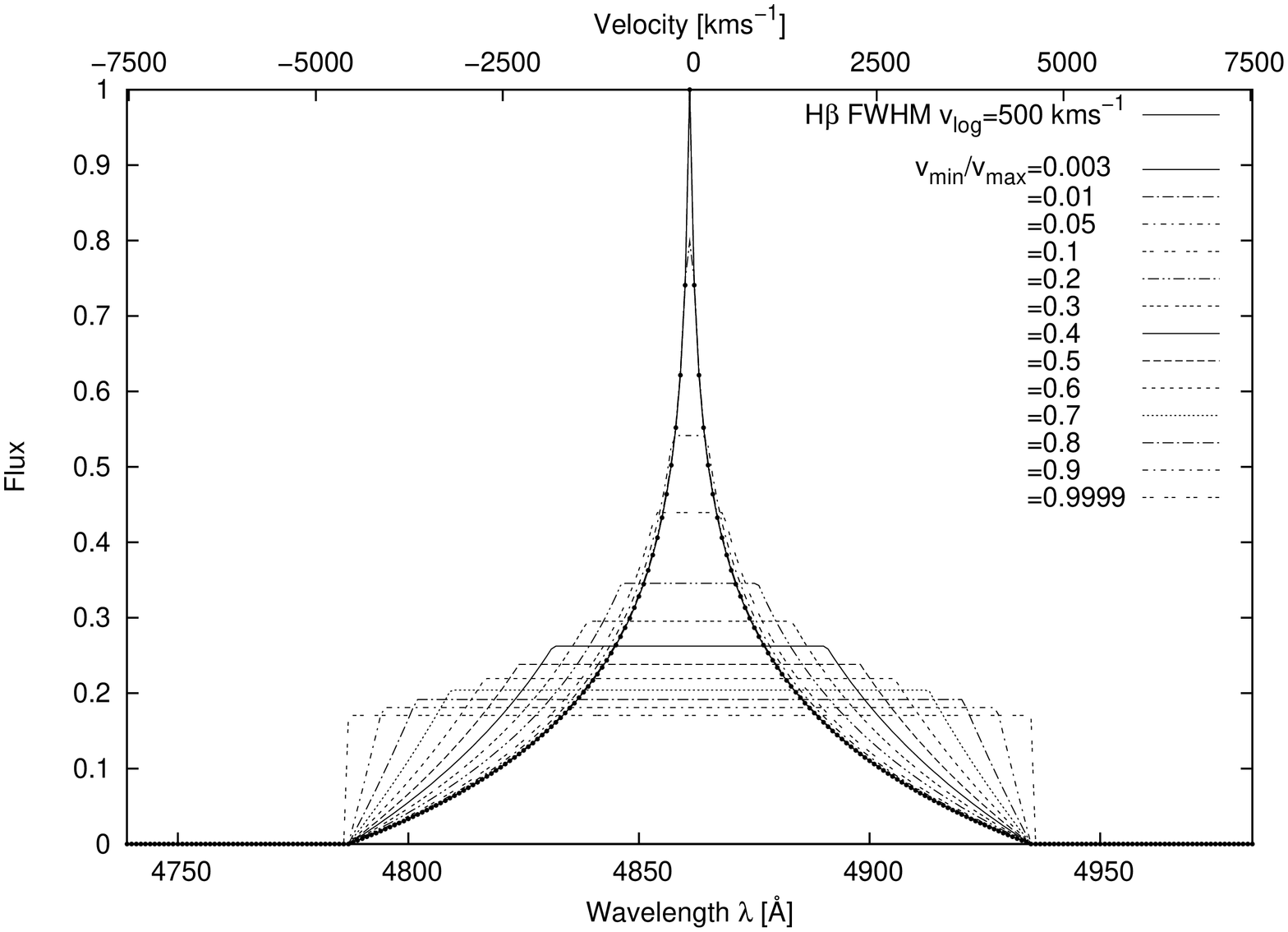} 
%\includegraphics[width=8.0cm,angle=0]{actdegrneigh.eps}
%\includegraphics[bb=40 90 380 700,width=63mm,angle=0]{actdegrneigh.eps}
%       \vspace*{5mm} 
%%       \vspace*{-2mm} 
  \caption{Logarithmic H$\beta$ profile with an intrinsic line width (FWHM)
 corresponding to 500\,$km\,s^{-1}$. Furthermore, we calculated the emission
line profiles caused by
expanding spherical shells of radiatively-driven isotropically emitting
clouds for different values of  $v_{min}$ to $v_{max}$ 
(see also Capriotti et al. \cite{capriotti80}).
}
   \label{profile_hb_log.ps}
\end{minipage}
\end{figure}
%
%-----------------------------------------------------------------------------
%
%------------------------------------------------------------------------------
%
\begin{figure}
\begin{minipage}[t]{0.475\textwidth}
\includegraphics[width=5.5cm,angle=270]{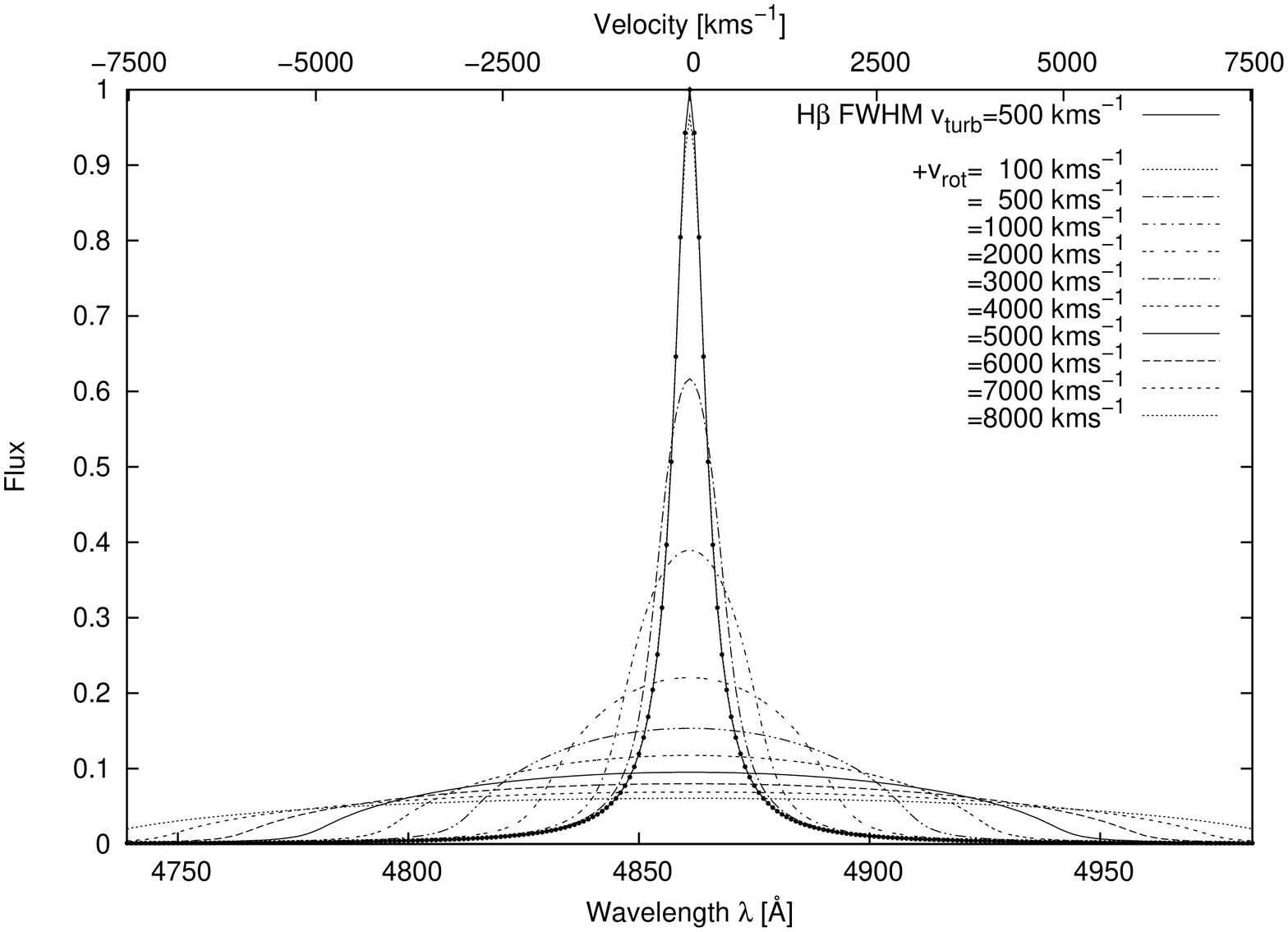} 
%\includegraphics[width=8.0cm,angle=0]{actdegrneigh.eps}
%\includegraphics[bb=40 90 380 700,width=63mm,angle=0]{actdegrneigh.eps}
%       \vspace*{5mm} 
%%       \vspace*{-2mm} 
  \caption{
Line broadening of a Lorentzian H$\beta$ profile due to rotation.
The rotation velocities range from 100\,$km\,s^{-1}$ to 8,000\,$km\,s^{-1}$.
The intrinsic turbulent velocity of the H$\beta$ line corresponds to
500\,$km\,s^{-1}$.
All profiles are normalized to have the same total flux.}
   \label{profil_examples.ps}
\end{minipage}
\hfill
\begin{minipage}[t]{0.475\textwidth}%
\includegraphics[width=5.5cm,angle=270]{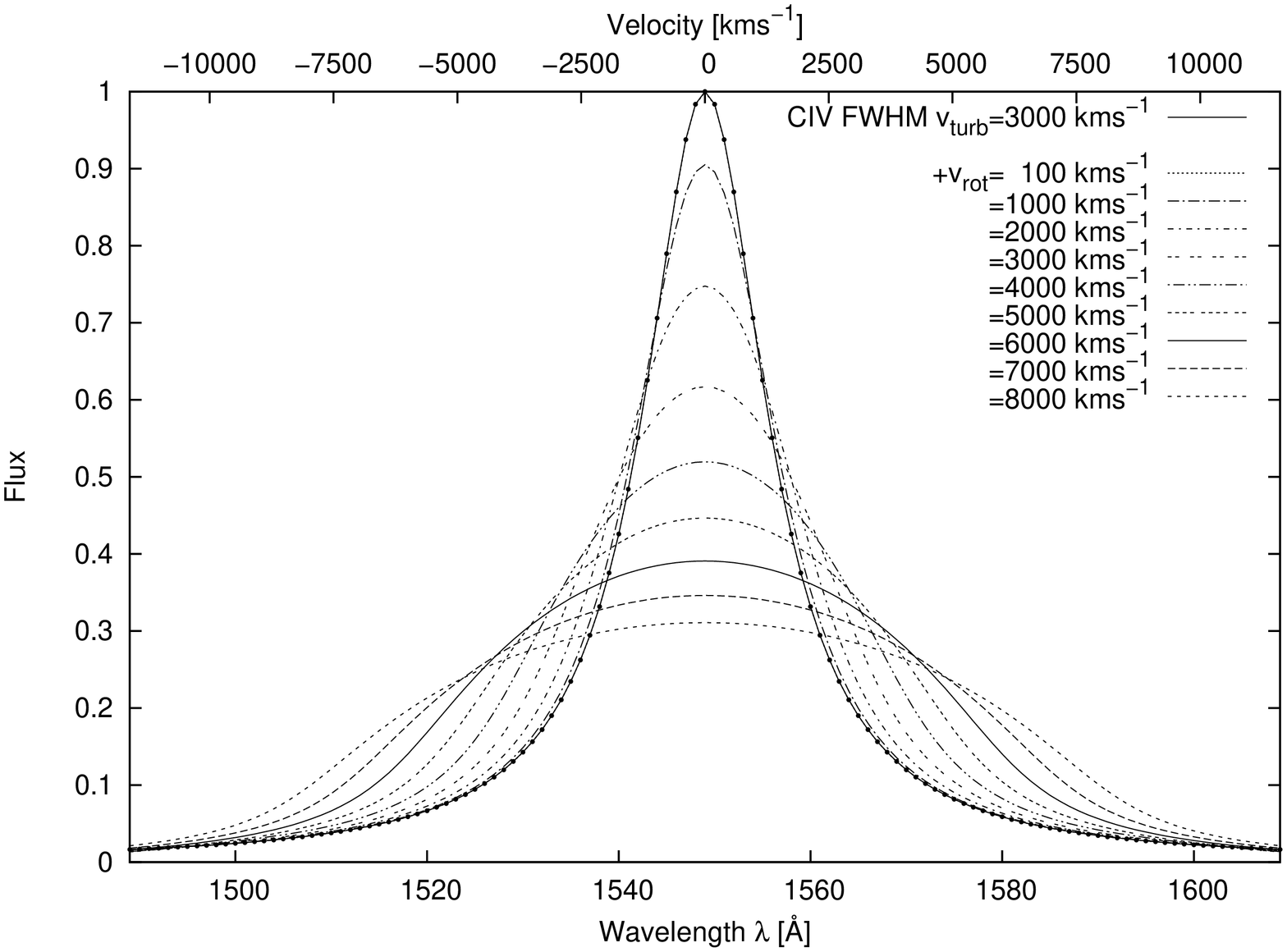} 
%\includegraphics[width=8.0cm,angle=0]{actdegrneigh.eps}
%\includegraphics[bb=40 90 380 700,width=63mm,angle=0]{actdegrneigh.eps}
%       \vspace*{5mm} 
%%       \vspace*{-2mm} 
  \caption{
Line broadening of a Lorentzian  \ion{C}{iv}\,$\lambda 1549$
profile due to rotation.
The rotation velocities range from 100\,$km\,s^{-1}$ up to 8,000\,$km\,s^{-1}$.
The intrinsic turbulent velocity of the \ion{C}{iv}\,$\lambda 1549$
line corresponds to 3,000\,$km\,s^{-1}$.
}
   \label{profil_examples_civ.ps}
%%\vspace{3.6mm} 
\end{minipage}
\end{figure}
%
%-----------------------------------------------------------------------------%
%------------------------------------------------------------------------------
%
\begin{figure}
\begin{minipage}[t]{0.475\textwidth}
\includegraphics[width=5.5cm,angle=270]{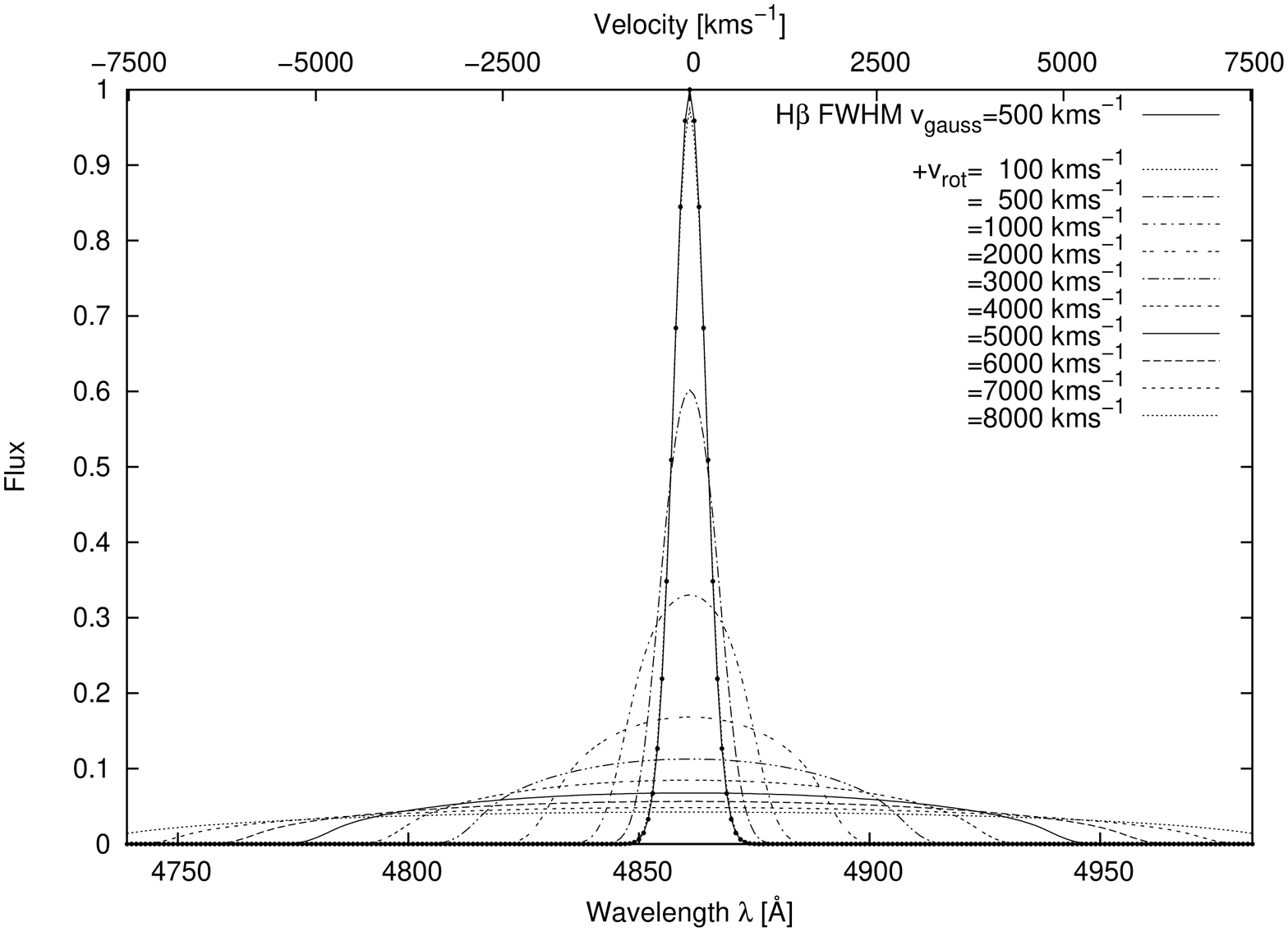} 
%\includegraphics[width=8.0cm,angle=0]{actdegrneigh.eps}
%\includegraphics[bb=40 90 380 700,width=63mm,angle=0]{actdegrneigh.eps}
%       \vspace*{5mm} 
%%       \vspace*{-2mm} 
  \caption{Line broadening of a Gaussian H$\beta$ profile due to
rotation. 
The rotation velocities range from 100\,$km\,s^{-1}$ up to 8,000\,$km\,s^{-1}$.
The intrinsic line width of the H$\beta$ line corresponds to
500\,$km\,s^{-1}$.}
   \label{profile_hb_gauss.ps}
%%\vspace{4.3mm} 
\end{minipage}
\hfill
\begin{minipage}[t]{0.475\textwidth}%
%%\vspace*{-2mm} 
\includegraphics[width=5.5cm,angle=270]{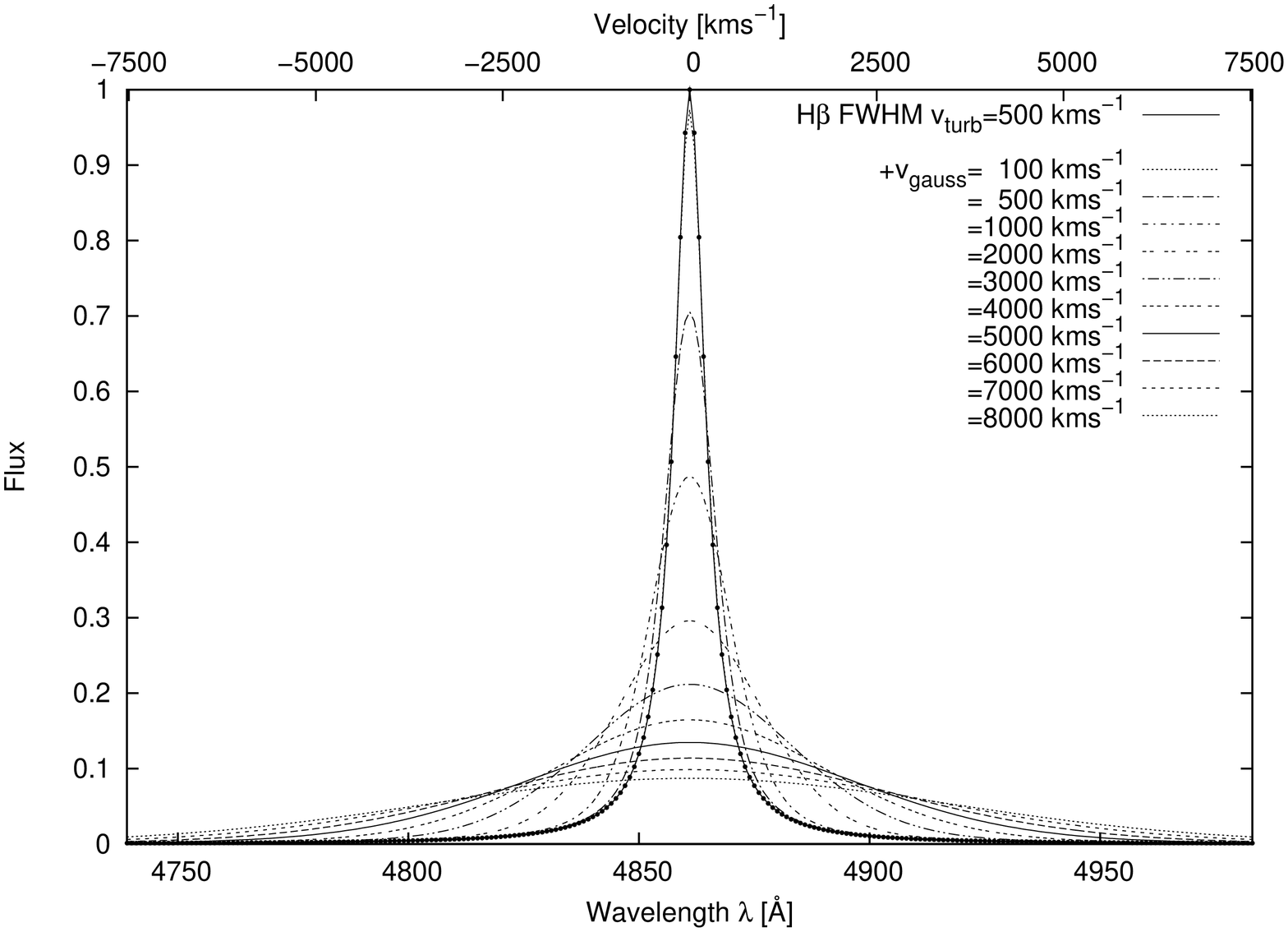} 
%\includegraphics[width=8.0cm,angle=0]{actdegrneigh.eps}
%\includegraphics[bb=40 90 380 700,width=63mm,angle=0]{actdegrneigh.eps}
%       \vspace*{5mm} 
%%       \vspace*{-2mm} 
  \caption{Line broadening of a Lorentzian H$\beta$ profile
due to Doppler motions.
The intrinsic line width of the H$\beta$ line corresponds to
500\,$km\,s^{-1}$. This Lorentzian profile has been convolved with
Gaussian profiles
having widths of 100\,$km\,s^{-1}$ up to 8,000\,$km\,s^{-1}$.}
   \label{profile_hb_lorentz_conv_gauss.ps}
\end{minipage}
\end{figure}
\twocolumn
%
%-----------------------------------------------------------------------------
%
%\clearpage
%
%
%\renewcommand{\textfraction}{0.3}
%
%
%------------------------------------------------------------------------------
%
\begin{figure}%[htbp]
\includegraphics[width=6.3cm,angle=270]{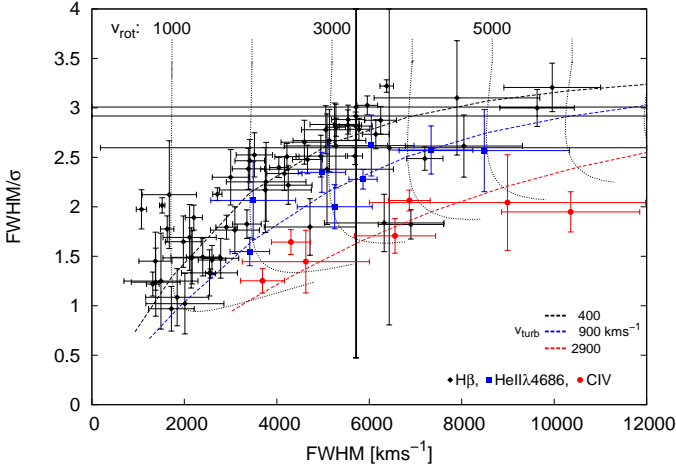} 
%\includegraphics[width=8.0cm,angle=0]{actdegrneigh.eps}
%\includegraphics[bb=40 90 380 700,width=63mm,angle=0]{actdegrneigh.eps}
%       \vspace*{5mm} 
%       \vspace*{-2mm} 
  \caption{Observed line width ratios
 FWHM/ $\sigma_{line}$ vs. line width FWHM for H$\beta$ (black diamonds),
\ion{He}{ii}\,$\lambda 4686$ (blue squares), and  \ion{C}{iv}\,$\lambda 1549$
(red circles). The dashed curves represent their corresponding theoretical
line width ratios based on rotational line broadened
 Lorentzian profiles (FWHM = 400, 900, and
2,900\,$km\,s^{-1}$). The rotation velocities go from 500 to 7000\,$km\,s^{-1}$
(curved dotted lines from left to right).  
 A color version of
this figure is available in the online journal.}
%\newline
   \label{hb_heii1640_civ.ps}
\vspace*{-4.8mm}
\end{figure}
%
%----------------------------------------------------------------------------- 
%
%------------------------------------------------------------------------------
%
\begin{figure}%[htbp]
\includegraphics[width=6.3cm,angle=270]{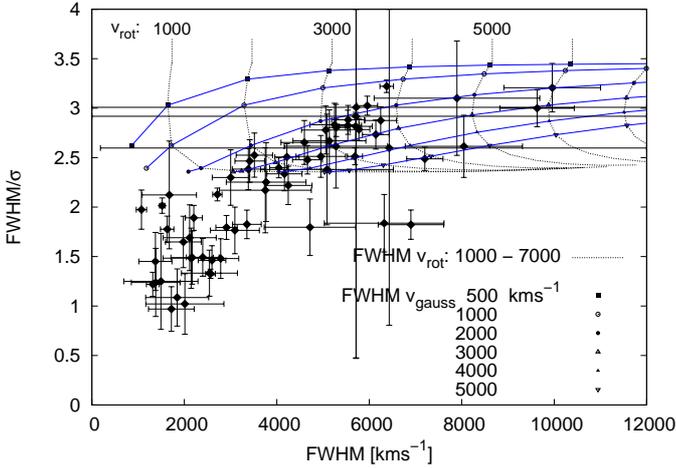} 
%\includegraphics[width=8.0cm,angle=0]{actdegrneigh.eps}
%\includegraphics[bb=40 90 380 700,width=63mm,angle=0]{actdegrneigh.eps}
%       \vspace*{4.5mm} 
%       \vspace*{-2mm} 
  \caption{Observed $H\beta$ line width ratios
 FWHM/ $\sigma_{line}$ vs. line width FWHM. The curves represent their 
modeling by rotational broadening
  (500 -- 7,000\,$km\,s^{-1}$)  of Gaussian profiles
(500, 1,000, 2,000, 3,000, 4,000, 5,000$km\,s^{-1}$
 line width). A color version of
this figure is available in the online journal.}
   \label{rot_gauss.ps}
\end{figure}
%
%----------------------------------------------------------------------------- 
%
%------------------------------------------------------------------------------
%
\begin{figure}%[htbp]
\includegraphics[width=6.3cm,angle=270]{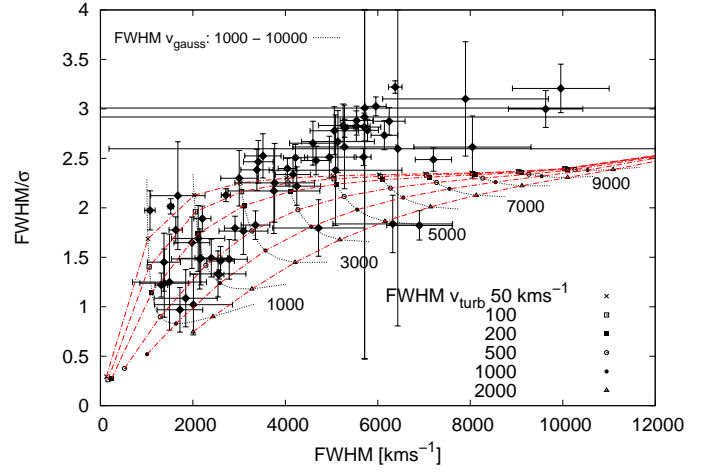} 
%\includegraphics[width=8.0cm,angle=0]{actdegrneigh.eps}
%\includegraphics[bb=40 90 380 700,width=63mm,angle=0]{actdegrneigh.eps}
%       \vspace*{5mm} 
%       \vspace*{-2mm} 
  \caption{Observed $H\beta$ line width ratios
 FWHM/ $\sigma_{line}$ vs. line width FWHM. 
The curves represent their modeling by convolving
Lorentzian profiles (50, 100, 200, 500, 1000, 2000\,$km\,s^{-1}$ line width)
with Gaussian profiles (500 -- 10,000\,$km\,s^{-1}$ line width). A color version of
this figure is available in the online journal.}
   \label{lorentz_conv_gauss.ps}
\vspace*{-2.9mm}
\end{figure}
%
%----------------------------------------------------------------------------- 
%
%
%------------------------------------------------------------------------------
%
\begin{figure}%[htbp]
\includegraphics[width=6.8cm,height=9cm,angle=270]{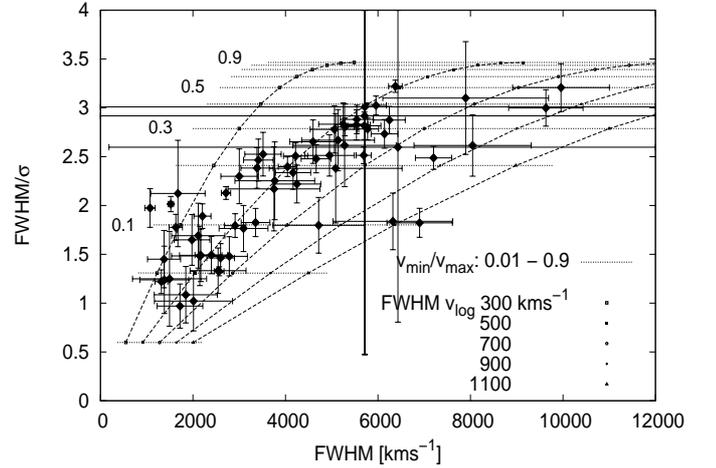} 
%\includegraphics[width=8.0cm,angle=0]{actdegrneigh.eps}
%\includegraphics[bb=40 90 380 700,width=63mm,angle=0]{actdegrneigh.eps}
%       \vspace*{5mm} 
%       \vspace*{-2mm} 
  \caption{Observed $H\beta$ line width ratios
 FWHM/ $\sigma_{line}$ vs. line width FWHM. The curves represent their 
modeling by logarithmic profiles
for different $v_{min}$ / $v_{max}$ ratios (0.01 - 0.9, from bottom to top).
Furthermore, we varied the line width from 300, 500, 700, 900, to 
1100\, $km\,s^{-1}$  (left to right).}
   \label{hb_log.ps}
\end{figure}
%
%----------------------------------------------------------------------------- 
%\clearpage
%

%------------------------------------------------------------------------------
%
\begin{figure}
\includegraphics[width=6.7cm,angle=270]{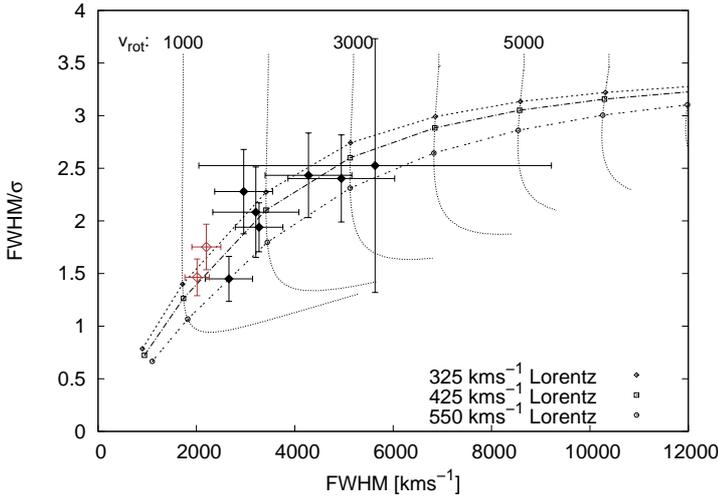} 
%\includegraphics[width=8.0cm,angle=0]{actdegrneigh.eps}
%\includegraphics[bb=40 90 380 700,width=63mm,angle=0]{actdegrneigh.eps}
%       \vspace*{5mm} 
  \caption{Observed $H\gamma$ line width ratios
 FWHM/$\sigma_{line}$ versus line width FWHM and their modeling by rotational
 broadening of Lorentzian profiles.
Less reliable measurements
are marked by open red diamonds.}
% \vspace*{-1.2mm} 
   \label{hg.ps}
\end{figure}
%----------------------------------------------------------------------------- 
%
%
%------------------------------------------------------------------------------
%
\begin{figure}
\includegraphics[width=6.7cm,angle=270]{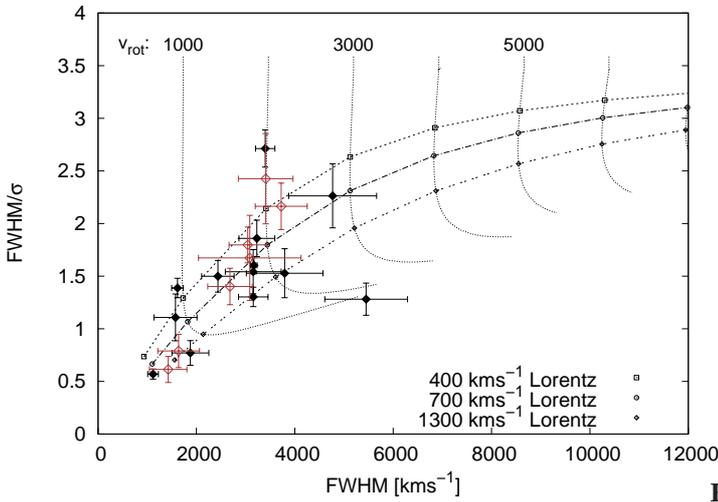} 
%\includegraphics[width=8.0cm,angle=0]{actdegrneigh.eps}
%\includegraphics[bb=40 90 380 700,width=63mm,angle=0]{actdegrneigh.eps}
%       \vspace*{5mm} 
%       \vspace*{-2mm} 
  \caption{Observed and modeled $H\alpha$ line width ratios
 FWHM/$\sigma_{line}$ versus line width FWHM.}
   \label{ha.ps}
\end{figure}
%
%----------------------------------------------------------------------------- 
%
%------------------------------------------------------------------------------
%
\begin{figure}
\includegraphics[width=6.7cm,angle=270]{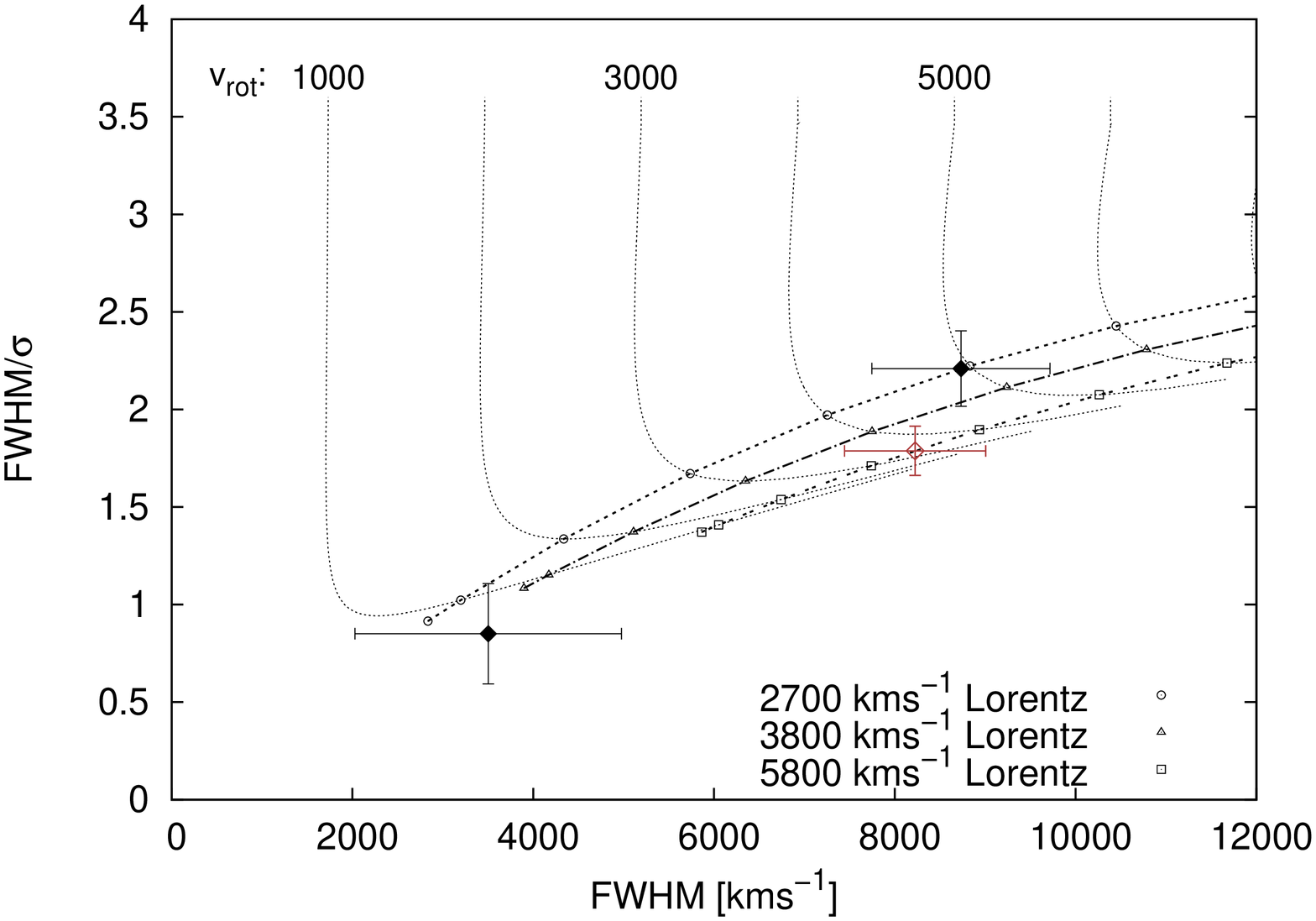} 
%\includegraphics[width=8.0cm,angle=0]{actdegrneigh.eps}
%\includegraphics[bb=40 90 380 700,width=63mm,angle=0]{actdegrneigh.eps}
%       \vspace*{5mm} 
%       \vspace*{-2mm} 
  \caption{Observed and modeled $Ly\alpha$+\ion{N}{v}\,$\lambda 1240$
 line width ratios
 FWHM/$\sigma_{line}$ versus line width FWHM.}
   \label{lya.ps}
\end{figure}
%
%----------------------------------------------------------------------------- 
%
%
%------------------------------------------------------------------------------
%
\begin{figure}
\includegraphics[width=6.7cm,angle=270]{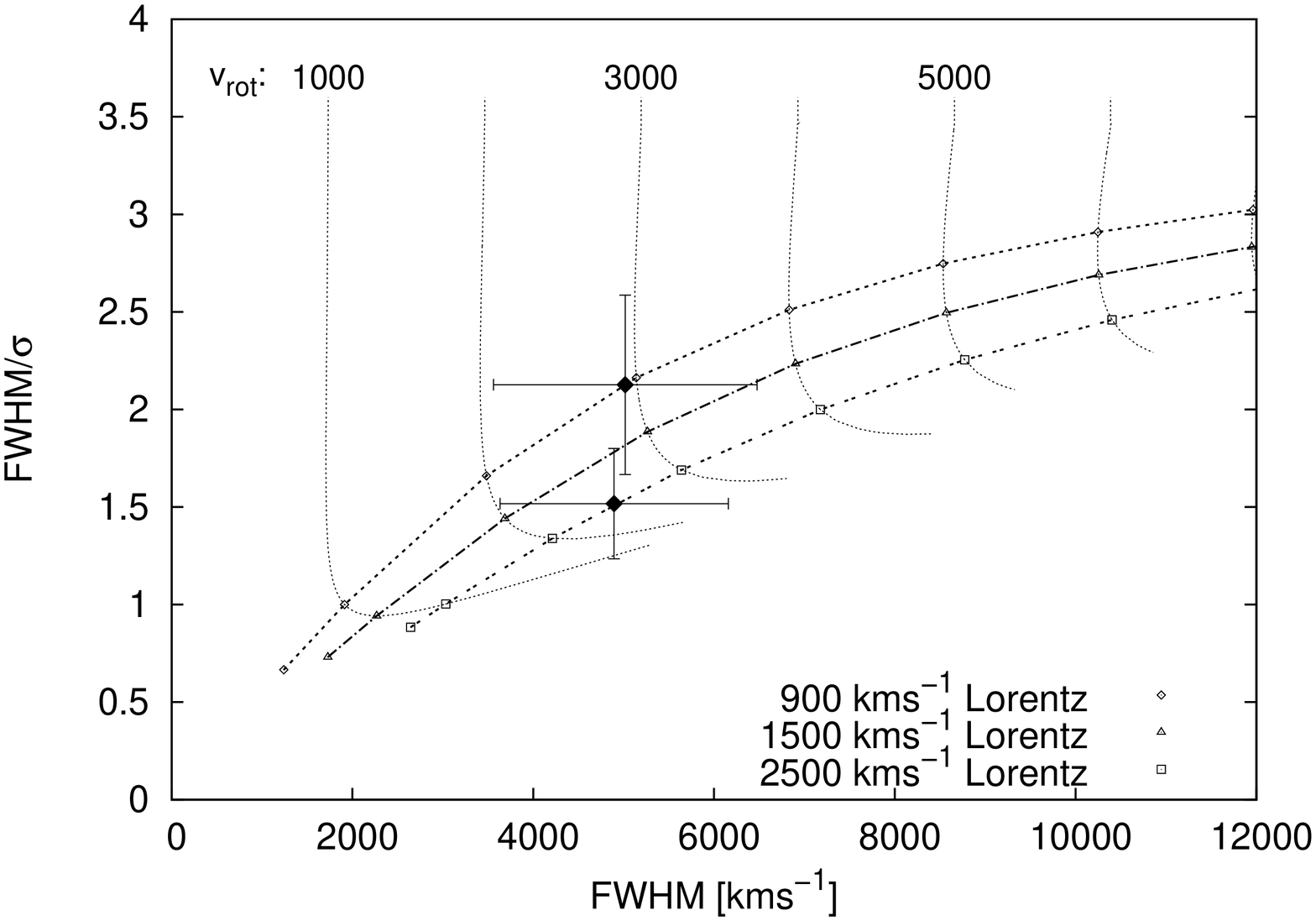} 
%\includegraphics[width=8.0cm,angle=0]{actdegrneigh.eps}
%\includegraphics[bb=40 90 380 700,width=63mm,angle=0]{actdegrneigh.eps}
%       \vspace*{5mm} 
%       \vspace*{-2mm} 
  \caption{Observed and modeled $\ion{C}{iii}]\,\lambda 1909$ line width ratios
 FWHM/$\sigma_{line}$ versus line width FWHM.\newline}
   \label{ciii1909.ps}
\end{figure}
%
%-----------------------------------------------------------------------------
%
%
%------------------------------------------------------------------------------
%
\begin{figure}
\includegraphics[width=6.7cm,angle=270]{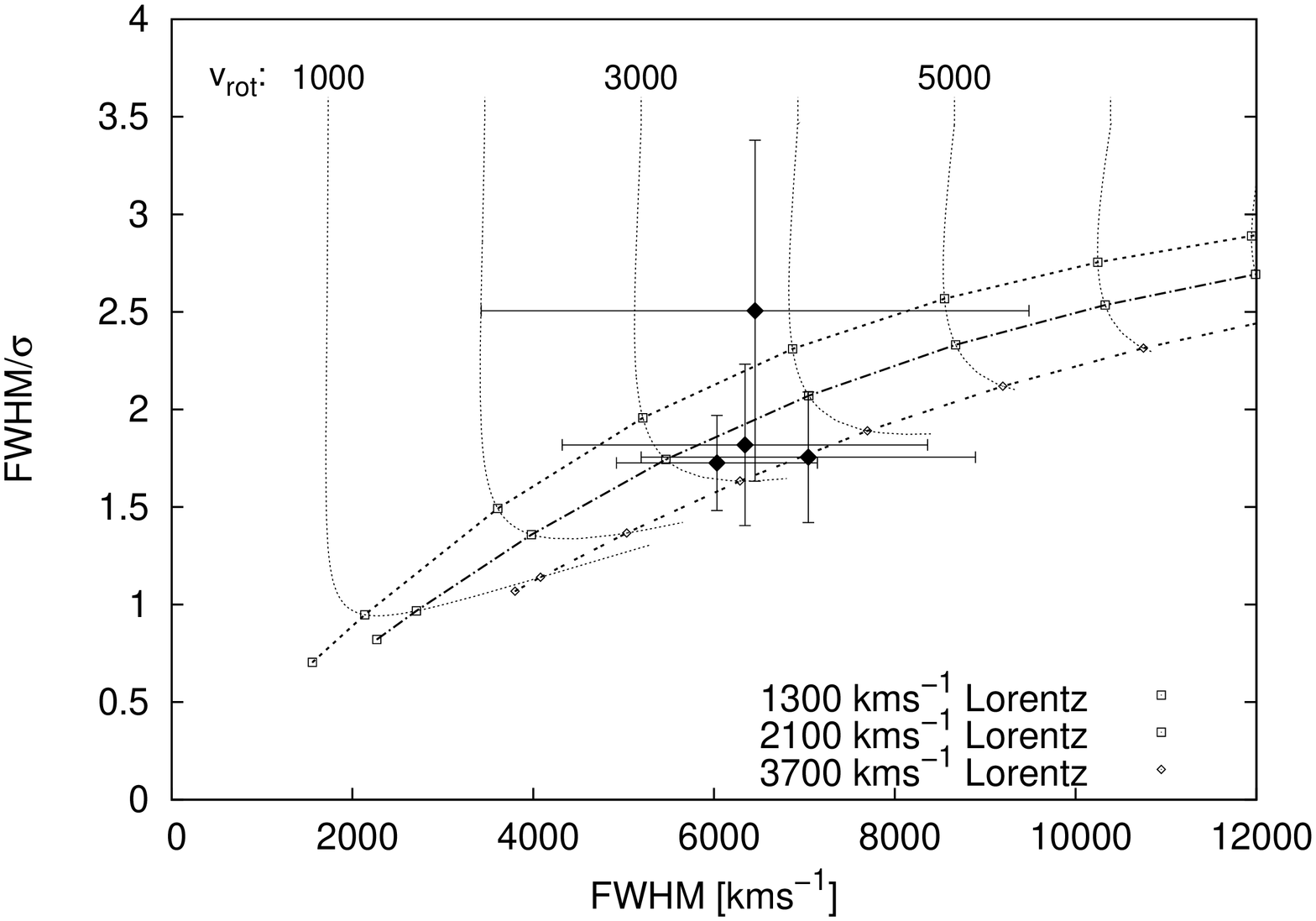} 
%\includegraphics[width=8.0cm,angle=0]{actdegrneigh.eps}
%\includegraphics[bb=40 90 380 700,width=63mm,angle=0]{actdegrneigh.eps}
%       \vspace*{5mm} 
%       \vspace*{-2mm} 
  \caption{Observed and modeled \ion{Si}{iv}\,$\lambda 1400$ line width ratios
 FWHM/$\sigma_{line}$ versus line width FWHM.}
   \label{siiv1400.ps}
\end{figure}
%
%-----------------------------------------------------------------------------
%
%------------------------------------------------------------------------------
%
\begin{figure}
\includegraphics[width=6.7cm,angle=270]{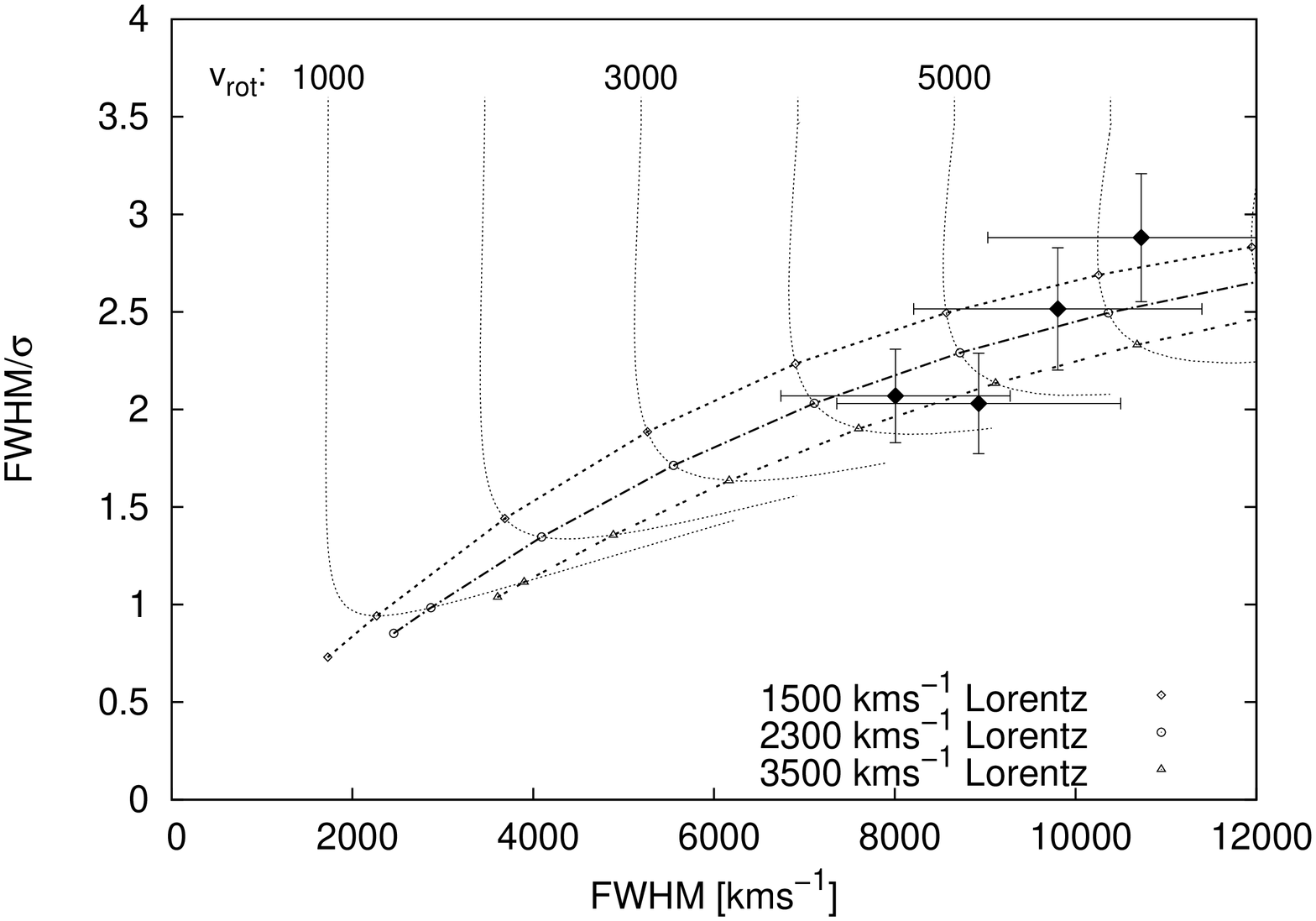} 
%\includegraphics[width=8.0cm,angle=0]{actdegrneigh.eps}
%\includegraphics[bb=40 90 380 700,width=63mm,angle=0]{actdegrneigh.eps}
%       \vspace*{5mm} 
%       \vspace*{-2mm} 
  \caption{Observed and modeled \ion{He}{ii}\,$\lambda 1640$ line width ratios
 FWHM/$\sigma_{line}$ versus line width FWHM.}
   \label{heii1640.ps}
\end{figure}
%
%-----------------------------------------------------------------------------
%\clearpage
%These errors are best fits by eye under the assumption that $\gtrsim$90
%percent of the data points should uniformly
% lie within the errors of the models
%over the whole FWHM range
%and that additional parameters influencing the line profiles are leading
%to smaller FWHM values than expected for their FWHM/ $\sigma_{line}$
%(see discussion section).
500 to 6,000\,$km\,s^{-1}$ in the individual
galaxies. The \ion{He}{ii}\,$\lambda 4686$ and           
 \ion{C}{iv}\,$\lambda 1549$ lines exhibit
 higher rotational velocities 
of at least 1,500\,$km\,s^{-1}$.
\subsubsection{Fitting the observed H$\beta$ line width ratios
with other models}
We generated further test models 
(Figs. 8 to 10) to adjust the observed H$\beta$
line width ratios by means of additional intrinsic line profiles and/or
further line broadening mechanisms. 
Figure 8 shows the computed trend of intrinsic
Gaussian profiles that are broadened by
rotation. The corresponding line profiles are presented in Fig.~5. 
The intrinsic Gaussian profiles
had line widths (FWHM) ranging from 500 to 5,000\,$km\,s^{-1}$
(from top to bottom).
The rotational velocities go from 500 to 
7,000\,$km\,s^{-1}$.
It is evident that these computed H$\beta$ line width ratios cannot explain the
observations because these broadened Gaussian profiles always exceed
FWHM/$\sigma_{line}$ ratio values of 2.35.
Lorentzian profiles, on the other hand, can easily explain low
FWHM/$\sigma_{line}$ ratios as seen in Fig.~7. 

Further line broadening tests of 
Lorentzian profiles convolved with Gaussian profiles that have
widths of 500 to 10,000\,$km\,s^{-1}$ explain neither the high observed
FWHM/$\sigma_{line}$ ratios nor the observed general trend of 
the FWHM/$\sigma_{line}$ to line width FWHM ratios (Fig.~9).
The corresponding line profiles
are given in Fig.~6. 

Figure~10 shows the trend of both observed and modeled $H\beta$ line
 width ratios
 FWHM/$\sigma_{line}$ versus FWHM for logarithmic profiles.
We modeled logarithmic profiles
for different $v_{min}$/$v_{max}$ ratios (0.01 - 0.9, from bottom
to top).
In addition we varied the line widths
 from 300, 500, 700, 900, to 
1100\, $km\,s^{-1}$  (left to right).
The corresponding line profiles are given in Fig.~2.

An exponential profile has a similar shape to a Lorentzian profile,
especially in their line wings (see Fig.~1). Line broadening due to rotation
of these two profile types leads to similar trends in the
FWHM/$\sigma_{line}$ vs. FWHM  figures. Exponential profiles have
a fixed  FWHM/$\sigma_{line}$ ratio of
FWHM/$\sigma_{line} = \sqrt{2}\ln{2} \approx 0.9803 $, while this ratio goes to
zero for Lorentzian profiles. 
All in all, the observed trend of varying line shape as a function of
line width - from the narrowest
 to the broadest line profiles -
cannot be explained by one varying parameter alone.

\subsubsection{Rotational line broadening of Lorentzian
 H$\gamma$, H$\alpha$,
Ly$\alpha$, \ion{C}{iii}]\,$\lambda 1909$, \ion{He}{ii}\,$\lambda 1640$,
and \ion{Si}{iv}\,$\lambda 1400$ profiles.}

All observed rms emission line width ratios
FWHM/$\sigma_{line}$ versus line width FWHM (from Table 1),
as well as their modeling,
are shown separately in Figs.~11 to 16
for the different emission lines H$\gamma$, H$\alpha$,
Ly$\alpha$, \ion{C}{iii}]\,$\lambda 1909$, \ion{He}{ii}\,$\lambda 1640$,
and \ion{Si}{iv}\,$\lambda 1400$.
The observed data are taken from the
AGN sample (see Peterson et al., \cite{peterson04}). 
Those measurements that
were regarded as less reliable by the authors
are marked by open red diamonds. These less reliable measurements follow
in most cases the general trend.
Besides the least square fits we show
lower and upper limits to the 
widths of the Lorentzian profiles (in Figs. 11 to 16).
 These limits are calculated from the variation in the individual data 
points with respect to the least square fit.

There are two H$\alpha$ measurements not following the trend:   
the galaxies NGC~4593 and PG~1426+015.
The galaxy NGC~4593 shows a high line width ratio
 FWHM/$\sigma_{line}$ of 2.7. This might be explained
with a high-inclination angle of the accretion disk (see Paper I)
in this galaxy. 
PG~1426+015 shows a low FWHM/$\sigma_{line}$ line width ratio with respect
to the derived line width of 5450\,$km\,s^{-1}$. The same galaxy was an outlier
in the corresponding  H$\beta$ figure (see Paper I). 

Furthermore, there is one \ion{Si}{iv}\,$\lambda 1400$ measurement (Fig.~15)
that does not follow the general trend by showing a high FWHM/$\sigma_{line}$
line width ratio. However, the error of this line measurement in the galaxy
NGC~5548 is very large.
A second measurement of this line
in NGC~5548 at FWHM/$\sigma_{line}$ = 1.755 (see Table 1) corresponds
 with the measurements of other galaxies.

The line width ratios
 FWHM/$\sigma_{line}$ versus line width FWHM of all these optical and UV
emission lines can be modeled in a simple way by rotational line
 broadening of Lorentzian 
profiles, as presented before in Paper I for the H$\beta$,
\ion{He}{ii}\,$\lambda 4686$, and \ion{C}{iv}\,$\lambda 1550$  lines.
 To each emission line belongs a dedicated
turbulent velocity that can be derived from the underlying Lorentzian profile.
This specific turbulent velocity of the individual lines
 is the same in all galaxies.
The intrinsic turbulent velocities belonging to the individual emission lines
are listed in Table~2. 

\begin{table}
\tabcolsep+2mm
 \newcolumntype{p}{D{p}{~~}{-1}}
\caption{Intrinsic turbulent velocities connected to the line emitting regions
of the strongest emission lines.} 
\centering
\begin{tabular}{lp}
\hline 
\noalign{\smallskip}
emission line  & \multicolumn{1}{c}{turbulent velocity} \\
 &   \multicolumn{1}{c}{[km\,s$^{-1}]$}   \\
%(1) & (2) & (3) & (4) \\ %& (5) & (6) & (7) & (8) & (9) \\ 
\noalign{\smallskip}
\hline 
\noalign{\smallskip}
H$\beta$       & 400p (-175)~(+300)\\
H$\gamma$      & 425p (-100)~(+125)\\
H$\alpha$      & 700p (-300)~(+600)\\
\ion{He}{ii}\,$\lambda 4686$  & 900p (-300)~(+200)\\
\ion{C}{iii}]\,$\lambda 1909$  & 1,500p (-600)~(+1,000)\\
\ion{Si}{iv}\,$\lambda 1400$  & 2,100p (-800)~(+1,600\\
\ion{He}{ii}\,$\lambda 1640$  & 2,300p (-800)~(+1.200)\\
\ion{C}{iv}\,$\lambda 1549$  & 2,900p (-1,000)~(+1,400)\\
Ly$\alpha$+\ion{N}{v}\,$\lambda1240$  & 3,800p (-1,100)~(+2,000)\\
\noalign{\smallskip}
\hline 
\end{tabular}
\end{table}

In Paper I we considered the possible effect of an inclination
of the accretion disk. An inclination leads to smaller line widths
in comparison to the intrinsic line widths. Here we did not correct 
for this in Table~2.

\section{Discussion}

\subsection{AGN line profiles}

In early studies of emission line profiles e.g.
Blumenthal \& Matthews (\cite{blumenthal75}) or
Capriotti et al. (\cite{capriotti80}) fitted
logarithmic profiles
to observed quasar emission lines.
 However, later on it was noticed that
their logarithmic fits result in line wings
that are too small in comparison 
to the observed profiles.
In those early years only very few high-quality
AGN spectra existed.

Afterwards it became apparent that the
profiles of narrow-line Seyfert 1 galaxies
with $H\beta$ line widths (FWHM) of less than 2000 - 4000\,$km\,s^{-1}$
are well adjusted by single Lorentzian profiles (Sulentic et al.
 \cite{sulentic00} and references therein,
Veron et al. \cite{veron01},
Marziani et al. \cite{marziani03}).
These authors confirmed previous claims that
observed broad Balmer line profiles are more precisely
fitted by Lorentzian rather than Gaussian profiles.
Gaussian profiles cannot reproduce the profiles observed
in narrow-line Seyfert 1 galaxies.

Laor (\cite{laor06}) found evidence
of exponential line wings 
in the H$\alpha$ line of the low luminosity narrow-line Seyfert 1
galaxy NGC~4395.
 However, this
galaxy is unique in having a
broad H$\alpha$ line width (FWHM) of $<$~520\,$km\,s^{-1}$ only. 
Laor (\cite{laor06}) explains this specific line profile
by electron scattering in the broad-line region.

Observed H$\beta$ profiles in AGN with line widths (FWHM)
 $<$ 4,000\,$km\,s^{-1}$ never can be explained by single Gaussian profiles
or by a combination of Gaussian profiles. Gaussian profiles have a constant
 FWHM/$\sigma_{line}$ ratio of 2.35. In contrast, all observed 
H$\beta$ profiles showing line widths (FWHM)
 $<$ 4,000\,$km\,s^{-1}$ exhibit lower FWHM/$\sigma_{line}$ ratios
(see Fig.~8). A combination of Lorentzian profiles with Gaussian profiles
does not match the observed trend in the
%line width ratio
 FWHM/$\sigma_{line}$ versus line width FWHM plots (see Fig.~9).
Logarithmic profiles meet the observed trend in the
FWHM/$\sigma_{line}$ versus FWHM plots
for those line profiles with line widths FWHM
 $<$ 6,000\,$km\,s^{-1}$ (Fig.~10). However, broader profiles
cannot be modeled in a simple way by varying only one parameter.

The profiles are not the same in all the quasars.
Sulentic et al. (\cite{sulentic00}) call those quasars
having H$\beta$ line widths (FWHM) $>$ 4,000\,$km\,s^{-1}$
Population~B quasars in contrast to Population~A quasars 
having H$\beta$ line widths (FWHM) $<$ 4,000\,$km\,s^{-1}$.
In a similar spirit, Collin et al. (\cite{collin06}) call those galaxies 
emitting line profiles narrower than Gaussian profiles
( FWHM/$\sigma_{line}$ $<$ 2.35) Population~1 AGN and those
galaxies showing line profiles broader than Gaussian profiles
Population~2 AGN (their Fig. 3).
Sulentic et al. (\cite{sulentic00}) modeled the broad H$\beta$ line profiles
(i.e. the Population~B quasars) with two Gaussian profiles, a broad one
and a narrower one. In no case could single Gaussian profiles match
their observed broad-line wings. 

We demonstrated in Paper I that the shape
(i.e. the FWHM/$\sigma_{line}$ ratio)
of the H$\beta$,  
\ion{He}{ii}\,$\lambda 4686$, and \ion{C}{iv}\,$\lambda 1550$ profiles 
varies systematically with their 
line width. There is no clear evidence for two separate quasar
populations.
 Here we show that this applies for
H$\gamma$, H$\alpha$, Ly$\alpha$, \ion{He}{ii}\,$\lambda 1640$,
\ion{C}{iii}]\,$\lambda 1909$, and
\ion{S}{iv}\,$\lambda 1400$ lines in our sample
 as well. The observed profiles of
all these emission lines can simply be
 characterized by rotational line broadening
of Lorentzian profiles. To every emission line belongs one exclusive
turbulent velocity of the line emitting region. The turbulent velocities
go from 400\,$km\,s^{-1}$ for the low-ionization lines up to
3,800\,$km\,s^{-1}$ for the high-ionization lines (see Table 2).
The rotational velocities
go from 500\,$km\,s^{-1}$ up to
6,500\,$km\,s^{-1}$.

Baskin \& Laor (\cite{baskin05}) claim
 that narrow CIV lines are rare
($\sim$2 per cent occurrence rate) compared with narrow
H$\beta$ $<$ 2,000\,$km\,s^{-1}$ ($\sim$20 per cent) based on more than
80 spectra from the Boroson-Green sample. 
This can be understood easily in terms of the high turbulent velocities
belonging to the \ion{C}{iv}\,$\lambda 1550$ line emission region
(2,900\,$km\,s^{-1}$) in comparison to the  H$\beta$ line region
(400\,$km\,s^{-1}$).

Line broadening due to turbulence and rotation are the main constituents
for the observed broad-line profiles in AGN.  
However, inclined accretion disk geometries of the line emitting regions
lead to smaller line widths owing to projection effects, while their
FWHM/$\sigma_{line}$ ratio remains constant (see Paper I).
Additional asymmetries in the line profiles might be caused by
geometrical/optical obscuration effects, additional outflow/inflow
components, anisotropic emission, superposition of line emission from
different emitting region, etc.  It was mentioned in Paper I that line
asymmetries lead to lower FWHM/$\sigma_{line}$ ratios as well.

 This general topic should
be studied in more detail in the future.    
Furthermore, obscuration affects 
individual line profiles in a different way 
because the lines originate in different regions
(see below). 

\subsection{Geometry and structure of the line emitting region}

In Table~2 one can identify the clear
 trend for higher ionized lines to
originate in those regions where higher turbulent velocities
are predominant.
As higher ionized lines exhibit broader emission lines in general
and usually originate closer to the center
(e.g. Peterson \& Wandel \cite{peterson99}, Kollatschny et al., 
\cite{kollatschny01}). This trend is consistent with a general increase
in the turbulent velocity towards central regions. 
The broad emission lines in AGN originate at distances of less than one light
day to more than 100 light days from the central ionizing source
 (e.g. Desroches et al. \cite{desroches06},
Kaspi et al. \cite{kaspi07}, Bentz et al. \cite{bentz09}).

Based on the earlier theoretical studies of Pringle (\cite{pringle81},
Eq. 3.16),
we made the following
claim about the geometry of AGN accretion disks in Paper~I.
The ratio of the accretion
disk height H with respect to their radius R is proportional to the 
turbulence velocity $v_{turb}$  in the accretion disk with respect to the
rotational velocity $v_{rot}$

\begin{equation}
\label{eq:HtoR}
  H/R = (1/\alpha) (v_{turb}/v_{rot}). 
\end{equation}

The unknown viscosity parameter $\alpha$ is assumed to be constant. 
We did not consider magnetic fields in this picture and
made the obvious assumption that the turbulence velocity $v_{turb}$
is less than the sound of speed $c_{S}$ in the disk:
  \begin{align}
    \label{eq:alp1}
    v_{turb} &< c_{S}\\ 
  v_{turb} &= \alpha'~c_{S}
  \end{align}

where $\alpha'$ is close to the usual $\alpha$ parameter.

The rotational velocities
- belonging to the individual emission lines - vary by a factor
of more than ten,
while the turbulent velocities that are
 connected with the individual emission lines
remain constant. This can be seen in Fig.~7, as well as in Figs.~11 to 16.
In this accretion disk model and accepting our results breaking
the velocity degeneracies one would conclude that slow-rotating
 AGN host a thick accretion disk
that is ten times thicker than fast-rotating AGN. 
A schematic picture of thick and thin accretion disks is shown in Fig.~17.
The radius of the central black hole ($r = 5.9\times10^{12} cm$) in Fig.~17
corresponds to a Schwarzschild mass of $M = 2\times10^{7} M_{\odot}$.

%------------------------------------------------------------------------------
%
\begin{figure}

\includegraphics[width=8.0cm,angle=0]{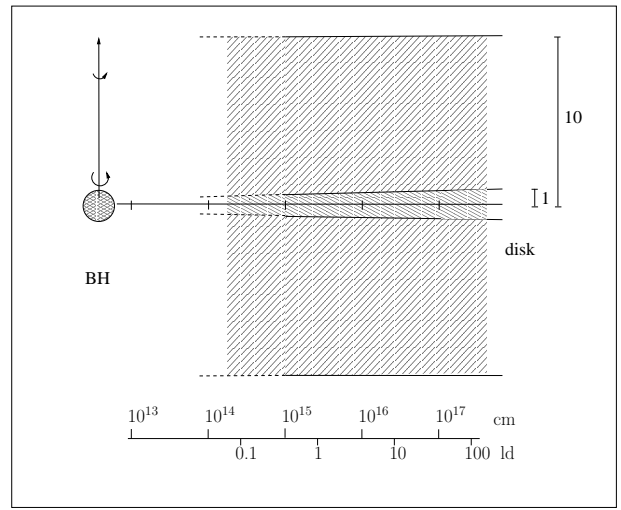} 

       \vspace*{5mm} 
%       \vspace*{-2mm} 
  \caption{
Schematic accretion disk models of AGN showing that slow-rotating AGN 
have an accretion disk thst is ten times thicker than fast-rotating AGN.}
   \label{blr_rot_model_names.eps}
\end{figure}
%
%-----------------------------------------------------------------------------

Other details regarding the physical conditions in
 the line emitting region seem to be 
even more complex: It has been noticed by means of
reverberation measurements that the \ion{He}{ii}\,$\lambda 4686$ and
\ion{He}{ii}\,$\lambda 1640$ lines in AGN spectra originate
at different distances from the center
(e.g. Peterson et al., \cite{peterson04}),
although they hold the same ionization degree.
Diverse models do not reproduce this observational fact
(e.g. Bottorff \cite{bottorff02}). 
Now we independently confirm 
the former finding
that these lines originate in different physical regions
based on the distinct turbulent velocities
we deduced for the two helium II line emission regions.
This is important considering that the
\ion{He}{ii}\,$\lambda 4686$/\ion{He}{ii}\,$\lambda 1640$ ratio
has sometimes been used as a reddening indicator for the broad-line region
(e.g. Snijders et al. \cite{snijders86}, Ferguson et al. \cite{ferguson95}).

The eigenvector studies of Boroson \& Green (\cite{boroson92})
demonstrated a strong correlation between the emission line width, optical
FeII emission, and soft X-ray photon index in AGN (Eigenvector 1).
The line width is connected with the geometry or rather, to be more specific,
with the thickness of the accretion disk (see above).
Therefore, the observed strength of the spectral
FeII emission and the X-ray photon index might be affected by geometry effects
as well.

\subsection{Correction factors for calculating central black hole
masses}

The central black hole mass $M_{BH}$ in AGN can be derived from
the broad emission line widths under the assumption 
that the gas dynamics are dominated by the central massive object:
\begin{equation}
\label{eq:MBH}
M_{BH} = f\,c\,\tau_{cent}\,\Delta\,v^{2}\, G^{-1}.
\end{equation}
 
%$M = f\,c\,\tau_{cent}\,\sigma^{2}\, G^{-1}  $.\\
%\[ M = \frac{f c}{G} v^{2} G^{-1} R . \]
%
The characteristic distances of the line-emitting regions
 $c\,\tau_{cent}$ from the central ionizing source
 can be derived by means of spectroscopic variability campaigns.
The distance can be calculated from the cross-correlation
function of emission line intensity variations with respect to
the ionizing continuum intensity
variations
 (e.g. Koratkar \& Gaskell \cite{koratkar91}, Kollatschny \& Dietrich
\cite{kollatschny97}).
The characteristic velocity $\Delta v$ of the emission line
region can be estimated from the FWHM of the rms profile or from
the line dispersion $\sigma_{line}$
(e.g. Peterson et al. \cite{peterson04}).
The scale factor f depends among others on the geometry and structure of the
line emitting region, as well as on their
inclination. Scale factors for rms spectra have 
numerical values from 0.5 up to 6.2 (e.g. Collin et al. \cite{collin06}),
depending on the assumptions
about geometry and structure of the line emitting region.
Only very few cases are known where the black hole mass could be estimated
independently by, e.g., gravitational redshift measurements
(Kollatschny \cite{kollatschny03b}) to get additional
 information on the scale factor f or rather the
inclination angle.

 Variability campaigns are very expensive with respect
to their observing time.
Kaspi et al. (\cite{kaspi05}) verify a relationship between AGN luminosity
and Balmer-line-averaged BLR size based on variability campaigns. This
scaling relation can be used
to obtain black hole masses based on single epoch spectra alone.
The velocity of the line emitting region can be estimated on
the basis of the width of the Balmer
emission lines. This method was extended to other emission lines,
especially to the \ion{C}{iv}\,$\lambda 1549$ line of distant high-redshift
AGN.

Central black hole masses were generally found to be bigger
for AGN showing broad H$\beta$ profiles 
(Pop. B), as well as broad \ion{C}{iv}\,$\lambda 1550$ profiles 
 than for those AGN that exhibit narrow H$\beta$ profiles  
 (Pop. A) (Marziani et al. 
\cite{marziani03}, Peterson et al., \cite{peterson04}, Vestergaard
\cite{vestergaard04},
Zamfir et al. \cite{zamfir10}).

The broader emission lines usually originate
closer towards the center as has been noted before (e.g. 
Peterson et al. \cite{peterson04}, Kollatschny \cite{kollatschny03a}).
The
turbulence grows towards the inner zones as well. Therefore
the black hole mass estimates, based on the line widths, are more heavily
biased in broad emission line objects caused by the additional turbulence
broadening
than in the narrower emission line objects.

We consider here nearby AGN where we have good black hole mass 
estimations from reverberation mapping.
Shen et al. (\cite{shen08}) have investigated biases
in virial black hole masses based on SDSS spectra.
Wang et al. (\cite{wang11}) discuss the influence of radiation-driven
outflows on CIV emission line profiles in high-redshift
and high-luminosity AGNs.
In the case of such outflows, the CIV emission lines should become broader
because of the additional line broadening component of the outflow.
If these
CIV profiles of high-redshift and high-luminosity AGNs show the same widths as
those of nearby AGNs, then the contribution
of the rotation on the line width must be even lower. In that case the
black hole mass estimation that is only based on the rotation velocity 
must be scaled down even more.

Vestergaard (\cite{vestergaard04}) derived black hole masses of nearby quasars 
based on their H$\beta$ line widths, as well as black hole masses
of high-redshift quasars based on their
\ion{C}{iv}\,$\lambda 1549$ line widths. 
%
%------------------------------------------------------------------------------
%
\begin{figure}
\includegraphics[width=6.3cm,angle=270]{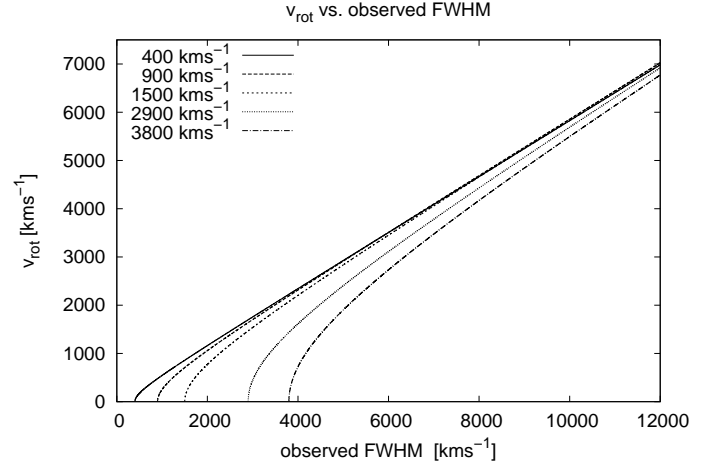} 
       \vspace*{5mm} 
%       \vspace*{-2mm} 
  \caption{
Relation between observed FWHM of the emission line profiles
and the related rotational velocity $v_{rot}$. The relation is shown for
intrinsic turbulence profiles with $v_{turb}$
ranging from 400\,$km\,s^{-1}$  up to 3,800\,$km\,s^{-1}$.}
   \label{intr_vs_obs_fwhm.ps}
\end{figure}
%
%-----------------------------------------------------------------------------
%
%------------------------------------------------------------------------------
%
\begin{figure}
\includegraphics[width=6.3cm,angle=270]{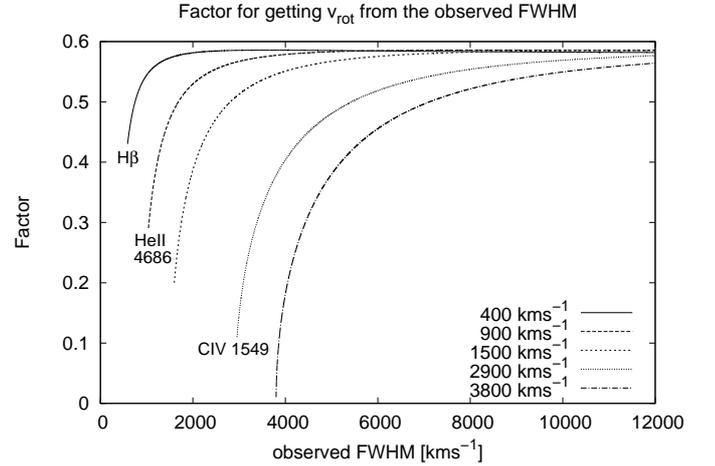} 
       \vspace*{5mm} 
%       \vspace*{-2mm} 
  \caption{
Factor for getting the rotational velocity $v_{rot}$ of
the line emitting region from
the observed FWHM, shown for observed profiles having intrinsic
turbulence velocities ranging from
 400\,$km\,s^{-1}$ up to 3,800\,$km\,s^{-1}$.}
   \label{factor_obs_intr_fwhm.ps}
\end{figure}
%
%--------------------------------------------------------------------
On average, she found bigger 
black hole masses - by a factor five to ten -
 for the younger high-redshift quasars based on their 
\ion{C}{iv}\,$\lambda 1549$ line widths in comparison to the old
nearby quasars based on their H$\beta$ line widths.  
In an additional paper,
Vestergaard \& Peterson (\cite{vestergaard06})
 published different scaling relations for nearby and
distant AGN black hole masses based on their H$\beta$, as well as
\ion{C}{iv}\,$\lambda 1549$ line widths.
However, Netzer at el. (\cite{netzer07}) note that
using the \ion{C}{iv}\,$\lambda 1549$ line width
 for estimating black hole masses
gives considerably different
results and a larger scatter than using the H$\beta$ line.
We demonstrated in Paper~I 
that black hole mass estimates based on the line width of the
\ion{C}{iv}\,$\lambda 1549$ line are overestimated by a factor of five to ten
in comparison to those black hole masses based on the width of the
 H$\beta$ line.  

On the basis of our observed and modeled line width ratios in Figs.~7, 11 - 16,
we present in Fig.~18 the relation between the
observed FWHM of the emission line
profiles and the related rotational velocity $v_{rot}$.
This relation is shown for underlying turbulence profiles with $v_{turb}$
ranging from
 400\,$km\,s^{-1}$  to 3,800\,$km\,s^{-1}$.
In Fig.~19
we present the correction factor towards the rotational
velocity $v_{rot}$ of the
broad-line clouds to calculate the central black hole masses of AGNs
based on the observed line widths.

Computed black hole masses go with the square of the rotational
velocity (see Eq. \ref{eq:MBH}).
Therefore, the corrected intrinsic black hole masses should be lower
by a factor of two to more than ten with respect to the calculated black hole
masses that have not been
corrected for the turbulent velocity contribution.
This correction factor is different for the different emission lines
(depending on the underlying turbulent velocity of the individual lines) 
and also depends on the additional rotational velocity broadening.
There is not an exclusive correction factor for black hole mass estimates
based on the
\ion{C}{iv}\,$\lambda 1549$ or H$\beta$ line widths as proposed by
Vestergaard \& Peterson (\cite{vestergaard06}) in their
mass scaling relations. Furthermore, black hole masses of distant
AGN based on the \ion{C}{iv}\,$\lambda 1549$ line widths have been
overestimated in comparison to nearby AGN where the mass estimate
is based on the H$\beta$ line. It is important
to consider this effect 
for our general understanding of the
evolutionary history of black hole masses in AGN.

\section{Conclusion}

We investigated the profile shapes of the UV/optical broad
emission lines in AGN in detail.
The two basic components causing
the line profile are Lorentzian profiles and rotational broadening.
An intrinsic turbulent velocity belongs
to each specific AGN emission line, which
 causes different line widths (FWHM) of the
particular Lorentzian profiles.
The turbulent velocities go from 400\,$km\,s^{-1}$ for H$\beta$ up to 
 3,800\,$km\,s^{-1}$ for Ly$\alpha$+\ion{N}{v}\,$\lambda 1240$.
The rotation velocities causing the line profile broadening 
 go from 500\,$km\,s^{-1}$  up to
 6,500\,$km\,s^{-1}$. There are not two separated
classes of broad-line AGN 
(narrow line and broad-line objects), but instead a continuous transition
from narrow to broad-line objects.
The accretion disk thickness in AGN can be derived from the ratio of the  
turbulent velocity $v_{turb}$ with respect to the
rotational velocity $v_{rot}$. Slow-rotating AGN 
have an accretion disk that is ten times thicker than fast-rotating AGN.
We find clear evidence that
 the  \ion{He}{ii}\,$\lambda 4686$ and
\ion{He}{ii}\,$\lambda 1640$ lines originate in different physical regions
in AGNs, although they hold the same ionization degree.
This finding is based on the distinct turbulent velocities
we deduce for the two helium II line emission regions.\\
Black hole mass corrections resulting from turbulent
velocity considerations are nontrivial and are larger for
\ion{C}{iv}\,$\lambda 1549$-based
measurements than H$\beta$, and both are a function of the observed FWHM.
In the literature
 usually the widths of the broad emission lines are used to
 compute the central black hole masses in AGN. However, one has to consider 
the contribution by the turbulence
 to get the corrected intrinsic rotational
 velocities. This investigation presented
 the individual correction factors towards the different emission lines 
for getting the intrinsic 
 rotational velocities $v_{rot}$ of
the line emitting region from
the observed FWHMs.
The corrected black hole masses are
lower than the uncorrected black hole masses 
 by a factor of two to more than ten,
depending on the emission lines, as well as on the rotational
velocities.
Especially those masses of the distant AGN
have been grossly overestimated 
where the masses have
been estimated on the basis of the \ion{C}{iv}\,$\lambda 1549$ line widths.

\begin{acknowledgements}
This work has been supported by the
Niedersachsen-Israel Research Cooperation Program ZN2318 and
DFG grant Ko 857/32-1.

\end{acknowledgements}

\newpage
%\begin{landscape}
%\setcounter{table}{3}
%\end{landscape}

\begin{thebibliography}{}
%
%\input{references}
 \bibitem[2005]{baskin05} Baskin, A. \& Laor, A. 2005,
MNRAS, 356, 1029
%
 \bibitem[2009]{bentz09} Bentz, M.~C. et al. 2009, ApJ, 705, 199
%
  \bibitem[1975]{blumenthal75} Blumenthal, G.,~R. \& Mathews, W.~G. 1975,
ApJ, 198, 517
%
 \bibitem[1992]{boroson92} Boroson, T. \& Green, R. 1992 ApJSuppl, 80, 109
%
 \bibitem[2002]{bottorff02} Bottorff, M.,~C. et al. 2002, ApJ, 581, 921
%
  \bibitem[1980]{capriotti80} Capriotti, E., Foltz, C. \& Byard, P. 1980
ApJ, 241, 903
%
 \bibitem[1988]{collin88} Collin-Souffrin, S., Dyson, J.~E., McDowell, J.~C.
\& Perry, J.~J. 1988, MNRAS, 232, 539
%
  \bibitem[2006]{collin06} Collin, S., Kawaguchi, T., Peterson, B.~M.,
 Vestergaard, M. 2006, A\&A, 456, 75
%
  \bibitem[2006]{desroches06} Desroches, L.-B. et al. 2006, ApJ, 650, 88
%
  \bibitem[2000]{elvis00} Elvis, M. 2000, ApJ, 545, 63
%
  \bibitem[1992]{emmering92} Emmering, R.T., Blandford, R.D., \& Shlosman, I.
       1992, ApJ, 385, 460
%
\bibitem[1995]{ferguson95} Ferguson, J.~W., Ferland, G.~J. \& Pradhan, A.~K.
               1995, ApJ 438, L55
%
%  \bibitem[2008]{gaskell08} Gaskell, C.~ M., Goosmann R.~W., Klimek, E.~S. 
% 2008, Mem SAI, 79, 1090
 \bibitem[2000]{fromerth00} Fromerth, M.~J. \& Melia, F.
               2000, ApJ 533, 172
%
\bibitem[2010]{gaskell10} Gaskell, C.~ M.  2010,
ASP Conf. Ser..  427,  68
%
 \bibitem[2012]{goad12} Goad, M.~R., Korista, K.~T. \& Ruff, A.~J. 2012,
    MNRAS, 426, 3086
%
 \bibitem[2008]{ho08} Ho, L. 2008, ARAA, 45, 475 
%
 \bibitem[2012]{hu12} Hu, C. et al. 2012, ApJ, in press (arXiv:1210.4187v1)
%
 \bibitem[1994]{hubeny94} Hubeny, I., Lanz,T. \& Jeffrey, C.S. 1994,
 in Newsletter of Astronomical Spectra No.20., ed. C.S. Jeffrey,
  St. Andrews Univ.  
%
%  \bibitem[1985]{hubeny85} Hubeny, I., Stefl, S. \& Harmonec, P. 1985,
%Bulletin of the Astronomical Institutes of Czechoslovakia  36,  214
%
  \bibitem[2005]{kaspi05} Kaspi, S. et al. 2005, ApJ, 629, 61
%
  \bibitem[2007]{kaspi07} Kaspi, S. et al. 2007, ApJ, 533, 631
%
  \bibitem[1994]{koenigl94} K\"{o}nigl, A., \& Kartje, J.E.
       1994, ApJ, 434, 446
%
 \bibitem[2001]{kollatschny01} Kollatschny, W. et al. 2001,
 A\&A, 379, 125
%
  \bibitem[2003a]{kollatschny03a} Kollatschny, W. 2003a, A\&A, 407, 461
%
  \bibitem[2003b]{kollatschny03b} Kollatschny, W. 2003b, A\&A, 412, L61
%
  \bibitem[1997]{kollatschny97} Kollatschny, W.,\& Dietrich, M. 1997,
       A\&A, 323, 5
%
  \bibitem[2011]{kollatschny11} Kollatschny, W. \& Zetzl, M. 2011,
Nature, 470, 366  (Paper I)
%
 \bibitem[1991]{koratkar91} Koratkar, A. \& Gaskell, M.  1991, ApJ 370, L61
%
  \bibitem[2006]{laor06} Laor, A. 2006, ApJ, 643, 112
%
  \bibitem[2003]{marziani03} Marziani, P. et al. 2003,
MNRAS, 345, 1133
%
 \bibitem[2012]{moriya12} Moriya, T.J. \& Tominaga, N.  2012, arXiv:1110.3807v2
%
 \bibitem[1997]{murray97} Murray, N., \& Chiang, J.  1997, ApJ, 474, 91
%
 \bibitem[1990]{netzer90} Netzer, H.  1990, 20 Saas-Fee Advanced Course of the
Swiss Society for Astrophysics and Astronomy: Active Galactic Nuclei (Berlin:
Springer), 57
%
  \bibitem[2007]{netzer07} Netzer, H. et al. 2007, ApJ, 671, 1256
%
  \bibitem[1978]{osterbrock78} Osterbrock, D.~E. 1978,
Proceedings of the National Academy of Sciences, 75,  540
%
 \bibitem[1999]{peterson99} Peterson, B.~M. \& Wandel, A. 1999,
ApJ, 521, L95 
%
 \bibitem[2000]{peterson00} Peterson, B.~M. \& Wandel, A. 2000,
ApJ, 540, L13 
%
 \bibitem[2004]{peterson04} Peterson, B.~M. et al. 2004,
ApJ, 613, 682
%
\bibitem[1981]{pringle81} Pringle, J.~E. 1981, ARAA, 19, 137
%
 \bibitem[2002]{richards02} Richards, G.~T. et al. 2002, AJ, 124, 1
%
\bibitem[2011]{richards11} Richards, G.~T. et al. 2011, AJ, 141, 167
%
  \bibitem[1995]{robinson95} Robinson, A. 1995,
MNRAS, 272, 647
%
 \bibitem[2008]{shen08} Shen, Y. et al. 2008, ApJ, 680, 169
%
 \bibitem[1986]{snijders86} Snijders, M.~H.~J., Netzer, H. \& Boksenberg, A.
              1986, MNRAS, 222, 549 
%
  \bibitem[2000]{sulentic00} Sulentic, J. W., Marziani, P. \&
Dultzin-Hacyan, D. 2000, Annual Review of Astronomy and
Astrophysics. 38, 521
%
  \bibitem[2002]{sulentic02} Sulentic, J. W. et al. 2002, ApJ, 566, L71
 %
  \bibitem[2012]{sulentic12} Sulentic, J. W., Marziani, P., Zamfir, S. \&
Meadows, Z.A. 2012, ApJ, 752, L7
%
  \bibitem[1955]{unsoeld55} Unsoeld, A. 1955, in 'Physics of stellar
   atmospheres', Springer publisher
%
  \bibitem[2001]{veron01} Veron-Cetty, M.-P., Veron, P., Goncalves, A.~C.,
 2001, A\&A, 372, 730
%
  \bibitem[2004]{vestergaard04} Vestergaard, M. 2004, ApJ, 601, 676
%
  \bibitem[2006]{vestergaard06} Vestergaard, M. \& Peterson, B.~M. 2006,
ApJ, 641, 689
%
  \bibitem[1995] {urry95} Urry, C.~M. \& Padovani, P. 1995, PASP 107, 803
%
  \bibitem[2011]{wang11} Wang, H. et al. 2011, ApJ, 738, 85
%
\bibitem[2010]{zamfir10} Zamfir, S., Sulentic, J. W., Marziani, P. \& 
Dultzin, D. 2010, MNRAS, 403, 1759
%
\end{thebibliography}
\end{document}